\documentclass[acmsmall,screen,nonacm,authorversion]{acmart}
\usepackage{listings}
\usepackage{enumitem}
\setlist{itemsep=3pt}
\usepackage{csquotes}
\usepackage[noabbrev,nameinlink]{cleveref}
\usepackage{comment}
\usepackage{xfrac}
\usepackage{hyperref}
\AtBeginDocument{%
  }

\setcopyright{none}

\def\LH/{Liquid Haskell}

\newcommand{\denotOf}[1]{\llbracket{#1}\rrbracket}
\newcommand{\remRef}[1]{\lfloor{#1}\rfloor}
\newcommand{\defsymb}{::=}
\newcommand{\evalContextSymb}{\mathbb{C}}
\newcommand{\evalContext}[1]{\evalContextSymb[#1]}
\newcommand{\lcName}{\lambda^\mathit{PC}}
\newcommand{\smallStep}{\hookrightarrow}

\newcommand{\triangleRight}{\scalebox{1.2}{\rotatebox[origin=c]{-90}{$\bigtriangleup$}}}

\definecolor[named]{constructor}{HTML}{009304}
\definecolor[named]{commentGray}{HTML}{4f4f54}

\newcommand\hsCommentStyle{\rmfamily\itshape\color{commentGray}}

\lstdefinelanguage{haskell}{
    string = [b]{"},
    morecomment = [l]{--},
    commentstyle = \hsCommentStyle,
    stringstyle = \color{blue},
	keywordstyle = [1]\color{constructor},
	keywordstyle = [2]\bfseries,
	keywordstyle = [3]\bfseries\color{red},
	keywords = [1]{Empty,Heap,Dist,Double,Int,Outcome,List,Nil,Cons,ProperDist,SubDist,NoDupList,IncrList,String,Node,Tree},
	keywords = [2]{data,type,Ord,Eq,reflect,measure,case,of,where,if,then,else,infix,let,in,import,lazy,QED},
	keywords = [3]{assume},
}

\def\Nat{\mathbb{N}}
\def\Int{\mathbb{Z}}
\def\Bool{\mathbb{B}}
\def\Real{\mathbb{R}}
\lstdefinestyle{haskell}{
    language=haskell,
    aboveskip=0.75\medskipamount,
    belowskip=0.75\medskipamount,
    numberstyle=\tiny\color{black},
    basicstyle=\color{black}\footnotesize\ttfamily,
    showstringspaces = false,
    basewidth={.5em,0.4em},
    escapechar=~,
    mathescape,
	keepspaces,
    xleftmargin=2em,
    captionpos=b,
	literate=
	  {->}{$\to$}{2}
      {<-}{$\gets$}{2}
	  {|>}{\triRight}{2}
	  {<=}{$\leqslant$}2
	  {>=}{$\geqslant$}2
	  {!=}{$\neq$}2
	  {=>}{$\Rightarrow$}2
      {<=>}{$\Leftrightarrow$}{2}
	  {\\}{$\lambda$}1
      {\#a}{$\alpha$}1
      {\#b}{$\beta$}1
      {\#c}{$\gamma$}1
      {\#N}{$\Nat$}1
      {\#Z}{$\Int$}1
      {\#B}{$\Bool$}1
      {\#R}{$\Real$}1
      {\#Rnneg}{$\Rnneg$}3
}
\NewDocumentCommand\C{ O{} }{\lstinline[style=haskell,basicstyle=\color{black}\footnotesize\ttfamily,#1]}
\NewDocumentCommand\FN{ O{#2} m O{black} }{\relax\ifmmode\hyperref[lst:#1]{\color{#3}\mathtt{#2}}\else\hyperref[lst:#1]{\color{#3}\footnotesize\texttt{#2}}\fi}

\def\listIn/{\FN[in]{\in}}

\def\meld/{\FN{meld}}

\newcommand{\heapSize}{}
\def\heapSize/{\FN[heapSize]{size}}

\newcommand{\bag}{}
\def\bag/{\FN{bag}}

\def\dunit/{\FN{dunit}}
\def\expectCost/{\FN{expectCost}}
\def\expectVal/{\FN{expectVal}}
\def\append/{\FN{append}}
\def\support/{\FN{support}}
\def\probOf/{\FN{probOf}}
\def\fMap/{\FN{fMap}}
\def\pMap/{\FN{pMap}}
\def\tick/{\FN{tick}}
\def\sumProb/{\FN{sumProb}}
\def\combine/{\FN{combine}}
\def\bind/{\FN{bind}}
\def\bernoulli/{\FN{bernoulli}}
\def\ProperDist/{\FN{ProperDist}[constructor]}
\def\SubDist/{\FN{SubDist}[constructor]}
\def\uniform/{\FN{uniform}}
\def\uniformBind/{\FN{uniformBind}}

\def\rquick/{\FN{rquick}}
\def\rquickBody/{\FN[rquick_body]{rquick\_body}}
\def\merge/{\FN{merge}}
\def\harmonic/{\texttt{harmonic}}

\def\Range/{\FN{Range}}

\def\thmSumFactor/{\FN[thm_sum_factor]{thm\_sum\_factor}}
\def\thmSumPermute/{\FN[thm_sum_permute]{thm\_sum\_permute}}
\def\thmSumLinear/{\FN[thm_sum_linear]{thm\_sum\_linear}}
\def\thmSumReverse/{\FN[thm_sum_reverse]{thm\_sum\_reverse}}
\def\thmSumConstant/{\FN[thm_sum_constant]{thm\_sum\_constant}}

\def\logTwo/{\FN[logTwo]{log2}}
\def\logTwoGeZero/{\FN[log2_ge_0]{log2\_ge\_0}}
\def\logTwoIneq/{\FN[log2_inequality]{log2\_inequality}}

\def\triRight{\hyperref[lst:triangle]{\color{black}$\triangleRight$}}

\newcommand\pow{}
\def\pow/{\FN{pow}}
\newcommand\partition{}
\def\partition/{\FN{partition}}

\NewDocumentCommand\Rnneg{}{%
	\mathbb{R}^{\geqslant 0}
}
\NewDocumentCommand\Eval{m m}{%
	\mathbb{E}_{#1}[#2]
}
\NewDocumentCommand\HEval{m m}{%
    \hyperref[lst:expectVal]{\color{black}\Eval{#1}{#2}}
}
\NewDocumentCommand\Ecost{m}{%
    \mathbb{E}^{\text{cost}}[#1]
}
\NewDocumentCommand\HEcost{m}{
    \hyperref[lst:expectCost]{\color{black}\Ecost{#1}}
}
\NewDocumentCommand\EcostOnly{}{\mathbb{E}^{\text{cost}}}
\NewDocumentCommand\HEcostOnly{}{\hyperref[lst:expectCost]{\color{black}\EcostOnly}}

\NewDocumentCommand\EcostOpen{}{\mathbb{E}^{\text{cost}}[}
\NewDocumentCommand\HEcostOpen{}{\hyperref[lst:expectCost]{\color{black}\EcostOpen}}

\NewDocumentCommand\HEvalOpen{ m }{\hyperref[lst:expectVal]{\color{black}\mathbb{E}_{#1}[}}

\NewDocumentCommand\finSum{ m m }{%
    \sum\hspace{-0.75em}\sum\nolimits_{\mathtt{#1}}^{\mathtt{#2}}
}
\NewDocumentCommand\HfinSum{ m m }{%
    \hyperref[lst:finSum]{\color{black}\finSum{#1}{#2}}
}

\DeclareMathOperator{\bO}{\mathsf{O}}
\newcommand*{\size}[1]{\lvert {#1} \rvert} 

\crefname{subsection}{subsection}{subsections}
\Crefname{subsection}{Subsection}{Subsections}

\begin{document}

%\title{Auto-Active Cost Analysis of Probabilistic Programs}
%\title{Zipping Through the Formalisation of Probabilistic Data Structures}
\title{To Zip Through the Cost Analysis of Probabilistic Programs}
% too funny? I'd prefer Data Structures, but 1-line title is better


\author{Matthias Hetzenberger}
\orcid{0000-0002-2052-8772}
\email{matthias.hetzenberger@tuwien.ac.at}
\affiliation{%
\institution{TU Wien}
\department[0]{Institute of Logic and Computation}
\department[1]{Research Unit for Formal Methods in Systems Engineering}
\city{Vienna}
\country{Austria}
}

\author{Georg Moser}
\orcid{0000-0001-9240-6128}
\email{georg.moser@uibk.ac.at}
\affiliation{%
\institution{University of Innsbruck}
\department{Department of Computer Science}
\city{Innsbruck}
\country{Austria}
}

\author{Florian Zuleger}
\orcid{0000-0003-1468-8398}
\email{florian.zuleger@tuwien.ac.at}
\affiliation{%
\institution{TU Wien}
\department[0]{Institute of Logic and Computation}
\department[1]{Research Unit for Formal Methods in Systems Engineering}
\city{Vienna}
\country{Austria}
}

\renewcommand{\shortauthors}{Matthias Hetzenberger, Georg Moser, and Florian Zuleger}

\begin{abstract}
Probabilistic programming and the formal analysis of probabilistic algorithms are active areas of research, driven by the widespread use of randomness to improve performance. While functional correctness has seen substantial progress, automated reasoning about expected runtime remains comparatively limited. In this work, we address this challenge by introducing a refinement-typed probability monad in \LH/.
Our monad enables automated reasoning about expected values and costs by encoding probabilistic behaviour directly in types. Initially defined for discrete distributions over finite support, it is extended to support infinite distributions via an axiomatic approach. By leveraging \LH/’s SMT-based refinement type checking, our framework provides a high degree of automation.
We evaluate our approach through four case studies: meldable heaps, coupon collector, randomised quicksort, and zip trees. The first two demonstrate automation with minimal annotation overhead. The latter two showcase how our monad integrates with interactive proofs, including the first formal verification of the expected runtime of zip trees.
\end{abstract}





\maketitle

%\listoftodos\relax

\section{Introduction}

\begin{table}
\caption{LOC of \LH/ Formalisations}
\label{fig:loc}

\begin{tabular}{|l|r|}
\hline
Probability Monad & 370 \\
Probability Monad Theorems & 2\,960 \\
Meldable Heaps & 122 \\
Coupon Collector's Problem & 27 \\
Quicksort: Functional Correctness & 117 \\
Quicksort: Expected Cost & 305 \\
Zip Trees: Functional Correctness & 591 \\
Zip Trees: Expected Cost & 7\,539 \\
Finite Sum Library & 602 \\
Logarithm Axiomatisation & 150 \\ \hline
\end{tabular}

\end{table}

Probabilistic programming and the automated analysis of probabilistic programs have become increasingly prominent in programming languages and formal verification research (see e.g.~\cite{BKS:2020,NipkowB19,EberlHN20,LMZ:2022,VasilenkoVB22,AvanziniBGMV24}). This growing interest is driven by the need to rigorously reason about systems that incorporate randomness, whether in decision-making, randomised algorithms, or data structures. Recent advances in semantics, verification frameworks, and type systems have enabled deeper insights into such systems. However, achieving high degrees of automation in this domain remains challenging due to the inherent complexity of probabilistic reasoning and the mathematical sophistication required to handle distributions, expected values and (expected)
cost models.

Probabilistic algorithms and data structures offer substantial performance benefits by leveraging randomness. 
The prototypical example of
the former is \emph{randomised quicksort}~\cite{Cormen:2009}, where the use of a random seed to choose the pivot uniformly yields the expected complexity of $\bO(n \log n)$ for a list
of length $n$. 
Further notable examples include \emph{coupon collector} schemes~\cite{MitzenMacherU05}, \emph{meldable heaps}~\cite{MartinezR98}, \emph{skip lists}~\cite{Pugh90}  and \emph{zip trees}~\cite{TarjanLT21}, all of which exhibit favourable expected runtime behaviour. While their theoretical efficiency is well established, the automated verification of their expected runtime has received comparatively less attention. This paper addresses this gap by presenting formal analyses of four prominent probabilistic case studies: meldable heaps, the coupon collector problem, randomised quicksort, and zip trees. These case studies serve as benchmarks to demonstrate our method’s applicability and scalability.

To facilitate the automated analysis of probabilistic cost, we propose a novel probability monad embedded in \LH/. Our approach builds on refinement types, a lightweight formalism that supports expressive specifications and leverages SMT solvers for automated verification. By encoding expected cost bounds and functional correctness directly in types, our probabilistic cost monad enables \LH/ to discharge complex probabilistic reasoning tasks automatically. This work advances prior efforts by extending \LH/’s capabilities to reason about probabilistic computations, thus narrowing the gap between manual proofs and fully automated verification.

We demonstrate that our monadic approach achieves a high degree of automation through detailed case studies on meldable heaps and the coupon collector problem. In these instances, the refinement types provide enough guidance for \LH/ to verify expected runtime bounds with minimal user annotations.%
\footnote{One may say ``we zip trough'' the formalisation effort in these case studies.}
Our results show that the ratio of proof annotations to program code remains low, indicating a substantial degree of automation in verifying probabilistic cost.

To further showcase the robustness of our approach, we extend our evaluation to randomised quicksort and zip trees. These examples highlight how our framework can combine automated refinement-type checking with more involved interactive proofs. Notably, our formalisation of zip trees is, to the best of our knowledge, the first complete formal verification of their expected runtime performance. Given that zip trees are functionally isomorphic to skip lists and share their desirable performance characteristics, this formal verification constitutes a significant milestone in the verification of modern probabilistic data structures.
In \Cref{fig:loc} we summarise the lines of code representing our formalisations, counting both Haskell code and refinement annotations.

\paragraph{Our contributions}

We present a general-purpose library -- encoded as a probability monad -- for implementing and verifying probabilistic algorithms in \LH/, supporting reasoning about both expected values and expected costs. Our contributions are as follows:

\begin{enumerate}
\item \emph{A refinement-typed probability monad.} We design a probability monad tailored for discrete distributions over finite support, leveraging refinement types to enable a high degree of automation. To extend its applicability, we later combine this monad with an axiomatic treatment of infinite structures, thereby supporting discrete distributions with infinite support (see \Cref{sec:zip-trees}).

\item \emph{Automation in probabilistic cost analysis.} We demonstrate the effectiveness of our approach through two case studies: meldable heaps and the coupon collector problem. In both cases, our system verifies expected runtime bounds with minimal annotation and proof effort, evidencing strong automation capabilities.

\item \emph{Integration with interactive verification.} We show that our probability monad integrates well with \LH/’s interactive proof features, enabling the verification of sophisticated probabilistic data structures. Notably, we present the first formal verification of the expected runtime of zip trees. Additionally, we provide a \LH/ library for reasoning about finite sums, which supports the verification of randomised quicksort and zip trees.
\end{enumerate}

% TODO
%\GM{table with LOCs of case studies}

\paragraph{Outline}

We start with motivating the central contribution of our work: a novel \emph{probability monad} for \LH/, allowing us to reason about expected costs and values simultaneously. Section~\ref{sec:automated-verification} provides
a gentle introduction in the use of refinement types for program verification, exemplified by establishing
a library for dealing with finite lists, as well as the axiomatisation of the logarithmic function.
The main technical contribution of this work is presented in \Cref{sec:prob-monad}, while
\Cref{sec:meldable-heaps} details the short case study on randomised meldable heaps. 
The Sections~\ref{sec:inf-dist} and~\ref{sec:zip-trees}
deal with the case study of the Coupon Collector's problem as well as with the novel formalisation
of zip trees. \Cref{sec:soundness} details soundness of our conservative extension of
\LH/. Finally, after discussing related work in \Cref{sec:related-work}, we
conclude in \Cref{Conclusion}.
In the Appendix, \Cref{sec:appendix}, we discuss the case study of randomised quicksort and
give the full proofs formalised in \LH/ for the theorems used in this paper.
Further, we make the implementation of the probability monad available as (online) Supplementary Material.

\newsavebox\lhBegin
\newsavebox\lhEnd
\begin{lrbox}{\lhBegin}
\lstinline[style=haskell,basicstyle=\color{black}\footnotesize\ttfamily]!{-@!
\end{lrbox}

\begin{lrbox}{\lhEnd}
\lstinline[style=haskell,basicstyle=\color{black}\footnotesize\ttfamily]!@-}!
\end{lrbox}

\section{Overview}
\label{sec:overview}

% GM: caption top or bottom?
\begin{figure}[t]
\centering
\begin{lstlisting}[style=haskell,numbers=left]
{-@
data Heap #a = Empty
            | Heap { key :: #a, left :: Heap {v:#a | key <= v}, right :: Heap {v:#a | key <= v} }

meld :: Ord #a => h1:Heap #a -> h2:Heap #a ~\label{lst:meld}~
     -> {result:ProperDist ({h:Heap #a | ~\bag/~ h = Bag_union (~\bag/~ h1) (~\bag/~ h2)}) ~\label{lst:meld:bag}~
	       | ~\expectCost/~ result <= ~\logTwo/~ (~\heapSize/~ h1) + ~\logTwo/~ (~\heapSize/~ h2)} ~\label{lst:meld:end-type}~
    / [~\heapSize/~ h1 + ~\heapSize/~ h2] ~\label{lst:meld:metric}~
@-}
meld Empty h2 = (~\logTwoGeZero/~ (~\heapSize/~ h2), ~\logTwoGeZero/~ (~\heapSize/~ Empty)) |> ~\dunit/~ h2
meld h1 Empty = (~\logTwoGeZero/~ (~\heapSize/~ h1), ~\logTwoGeZero/~ (~\heapSize/~ Empty)) |> ~\dunit/~ h1
meld (Heap k1 l1 r1) (Heap k2 l2 r2)
    | k1 <= k2 =
           ~\logTwoIneq/~ (~\heapSize/~ l1) (~\heapSize/~ r1) ~\label{lst:meld:logIneq}~
        |> ~\bernoulli/~ 0.5 ~\label{lst:meld:bernoulli}~
               (~\fMap/~ (~\tick/~ 1 (meld l1 (Heap k2 l2 r2))) (\ h -> Heap k1 h r1))
               (~\fMap/~ (~\tick/~ 1 (meld r1 (Heap k2 l2 r2))) (\ h -> Heap k1 l1 h))
    | k2 < k1 =
           ~\logTwoIneq/~ (~\heapSize/~ l2) (~\heapSize/~ r2)
        |> ~\bernoulli/~ 0.5
               (~\fMap/~ (~\tick/~ 1 (meld (Heap k1 l1 r1) l2)) (\ h -> Heap k2 h r2))
               (~\fMap/~ (~\tick/~ 1 (meld (Heap k1 l1 r1) r2)) (\ h -> Heap k2 l2 h))
\end{lstlisting}
\caption{A Probabilistic Data Structure: Meldable Heaps}
\label{lst:meldableheaps}
\end{figure}

We motivate the central contribution of our work: a novel \emph{probability monad} for \LH/, allowing us to reason about expected costs and values simultaneously.
First, we consider the core implementation of a simple probabilistic data structure, namely
\emph{randomised meldable heap}~\cite{GambinM98}. This aims at showcasing the support provided by our probability monad
in automatically establishing its expected cost analysis via a refinement type system.
Incidentally, the ease of this formalisation exemplifies our motivation to use a refinement type system like \LH/ as the basis of our work.
Second---using this motivational example---we provide a high-level overview of the established probability monad. Here, we highlight its core features, providing for the formal verification of the expected cost and value analysis of sophisticated probabilistic data structures.
Third, we briefly emphasise one of the major case studies, detailed in the following, namely the formalisation of the expected cost analysis of \emph{zip trees}~\cite{TarjanLT21}, a data structure isomorphic to \emph{skip lists}~\cite{Pugh90}.

\paragraph{Randomised meldable heaps}

Meldable heaps derive their name from their central operation of \emph{melding}
that suitably combines two heaps $h_1$ and $h_2$ into a new one. The innovation of
\emph{randomised meldable heaps} is to perform this operation subject to a fair coin toss, resulting
in an (expected) bound of the cost of the operation of $\log_2(\size{h_1}) + \log_2(\size{h_2})$.
More concretely, consider~\Cref{lst:meldableheaps}. Our encoding starts off with a
standard type definition of heaps in a refinement type setting.%
\footnote{We use the \LH/ notation {} \usebox\lhBegin {} and {} \usebox\lhEnd {} to define refinement types and refinement type specifications.}
Eliding its type signature for a moment, the function \meld/ implements
the melding functionality, making use of sampling from a Bernoulli distribution to represent
fair coin tosses.
This implementation is perfectly standard, with two exceptions. On the one hand the command
\tick/ allows us to seamlessly represent a (monotone) cost model. Each of the recursive calls
is accounted for with cost~$1$. On the other hand, the annotation
\C!$\logTwoIneq/$ (size l1) (size r1)! is required
to support the automated derivation of the optimal logarithmic upper bound on the expected cost.

Let us now consider the refinement type definition in lines \ref{lst:meld}--\ref{lst:meld:end-type} in~\Cref{lst:meldableheaps} of the function \meld/ more closely.
\begin{center}
\begin{minipage}{\linewidth}
\begin{lstlisting}[style=haskell]
{-@ meld :: Ord #a => h1:Heap #a -> h2:Heap #a ->
            {result:ProperDist ({h:Heap #a | bag h = Bag_union (bag h1) (bag h2)})
                   | expectCost result <= log2 (size h1) + log2 (size h2)} @-}
\end{lstlisting}
\end{minipage}
\end{center}  
This type definition demonstrates the introduced probability monad. It states that for any ordered type of elements $\alpha$, the melding of two heaps over such
elements results in a distribution of heaps of the union the elements in the two heaps. This
part represents the functional correctness of the algorithm. Further, the refinement
type also states the aforementioned cost bound on the expected cost of \meld/.
We have set up our probability monad, expressing expected costs (and values),
as well the ticking functionality in such a way that \LH/ automatically discharges
any of the constraints induced by the given refinement type. This showcases the power of
a refinement type system as it allows the automated verification of
the optimal bound on the (expected) cost of (randomised) meldable heaps.
In this sense, our probability monad provides the user of \LH/ with significantly increased
automation support for the analysis of a typical (probabilistic) data structure, cf.~\cite{Matheja20,LMZ:2022}.

\paragraph{Probability monad}

Following the literature we implement probability distributions (and our cost model)
as a \emph{probability monad}, following similar approaches in the literature,
cf.~\cite{Ramsey02, Erwig06, Scibior15, VasilenkoVB22}. In contrast to, for example~\cite{VasilenkoVB22},
however, we build up this monad from first principles,
rather than wrap it around an existing Haskell library supporting a monadic interpretation of probabilities

\begin{minipage}{\linewidth}
\begin{lstlisting}[style=haskell]
  type Dist #a = ... -- ~\hsCommentStyle defined in \Cref{sec:prob-monad} as list of probability outcomes~   

  {-@ type ProperDist #a = {d:Dist #a | ~\sumProb/~ d = 1} @-}
  {-@ type SubDist #a = {d:Dist #a | ~\sumProb/~ d <= 1} @-}
\end{lstlisting}
\end{minipage}\vspace{0.5\medskipamount}

Using (finite) distributions, we can now define (proper) \emph{distributions}, using the type
\C!ProperDist! above, as well as \emph{subdistributions} (via type \C!SubDist!).
These type definitions use refinements, in our case the condition that the probabilities either sum up to exactly one (for \ProperDist/) or at most one (for \SubDist/).
To exemplify the use of refinement
types in the formalisation of functional correctness arguments re-consider the motivating
example of randomised meldable heaps. In particular let us look afresh at the refinement type signature of \meld/ above.
As emphasised, two heaps $h_1,h_2$ over elements of (ordered) type $\alpha$ are melded in a
probabilistic fashion, yielding a (proper) distribution of heaps, containing the combined elements
of $h_1$ and $h_2$, respectively. Note, how the type \ProperDist/ expressed the correct
return type.

Slightly simplified, the expected cost and the expected value of a random variable~\C!f!
with respect to a distribution can now be given as follows, focusing again on the refinement type signature (see~\Cref{sec:prob-monad} for the full definition).
\begin{lstlisting}[style=haskell]
  {-@ expectCost :: Dist #a -> #Rnneg @-}
  {-@ expectVal :: (#a -> #R) -> Dist #a -> #R @-}
\end{lstlisting}

Kindly note how seamlessly the function \expectCost/ can be employed to express the expected cost
of melding, which can now be upper bound directly in the refined type signature. At this point
it is important to emphasise that our formalisation makes use of an axiomatic treatment of
logarithmic functions. This has turned out to allow for a very smooth representation.
We conclude our high-level description of our probability monad, by stating its core functionality.
\begin{lstlisting}[style=haskell]
  {-@ dunit :: f:(#a -> #R) -> x:#a -> 
               {r:ProperDist #a | ~\expectCost/~ r = 0 && ~\expectVal/~ f r = f x} @-}
  
  {-@ bind :: d:Dist #a -> (#a -> ~\ProperDist/~ #b) -> 
              {r:Dist #b | ~\sumProb/~ r = ~\sumProb/~ d} @-}
\end{lstlisting}

Injecting an arbitrary value into a dirac distribution uses the function \C!dunit x!. Note that the
refinement not only specifies the probabilities, but also the expected value, parametrised in a yet unknown random variable~\C!f!. On the other hand, the function \C!bind! enables the sequential composition of monadic computations over distributions.

\paragraph{Zip trees}

These binary search trees are a probabilistic data structure, recently invented by \citet{TarjanLT21}.
Zip trees are closely related to \emph{treaps}~\cite{AragonS89,BlellochR98} except ranks are generated through a geometric distribution and insertions and deletions are handled through unzipping and zipping rather than tree rotations. Further, zip trees are isomorphic to \emph{skip lists}~\cite{Pugh90}, but have a better (expected)
worst-case costs of $\bO(\log(n))$ for insertion and deletion  for a tree containing $n$~elements.
This improved speed in relation to skip lists, also explains their name. Paraphrasing from~\cite{TarjanLT21} ``to zip'' as something that moves really fast.

The pen-and-paper complexity proof established in~\citet{TarjanLT21} uses a very clever abstraction of the
problem, which does not immediately reveal a straightforward path toward formalisation. Indeed, to best of our knowledge, no formalisation of either formal correctness nor expected cost bound of an implementation of zip trees has been reported in the literature.
Employing our novel probability monad, we have established the first such formalisation of these properties.
Our formalisations naturally handles functional correctness of the implementation,
as well as the verification of the aforementioned cost bound with surprising ease. 
In contrast to the formalisation of randomised meldable heaps, however, our formalisation does require extrinsic proofs, as the constraints induced by the refinement type system can no longer automatically be discharged. Nevertheless, we managed to follow the original pen-and-paper proof quite closely, resulting in a compact and readable formalisation. Our formalisation yields the precise estimate on the expected cost of $\sfrac{3}{2} \cdot (\log_2(n) + 1)$, thereby slightly improving the pen-and-paper proof.

\section{Auto-Active Verification with \LH/}
\label{sec:automated-verification}

Refinement types allow associating types of a programming language with logical predicates with the goal of enforcing properties at compile time.
\LH/ implements refinement types for Haskell with a high degree of automation, by reducing type checking to verification constraints efficiently decidable SMT solvers.
Moreover, it is possible to reason about Haskell programs within Haskell, making it possible to formalise, verify, and --- to a certain extent --- automate pen-and-paper proofs.
\LH/ can be described as an \emph{auto-active} verification tool \cite{leino2010usable}.
Auto-active verifiers take both code and specifications before as input before verification constraint generation and proceed without interaction until verification succeeds or fails.
In contrast, an interactive verifier often needs guidance from the user during the verification process.
In this section we will give a short introduction to verification and show two ways of performing mathematical proofs with \LH/.
As \LH/ encodes Haskell's \C!Int! and \C!Double! types to unbounded integers and real numbers as supported by the SMT solver, we will write \C!$\mathbb{Z}$! for \C!Int! and \C!$\mathbb{R}$! for \C!Double!.

\subsection{Reasoning about Programs}
\label{subsec:reasoning-about-programs}

Let us consider simple lists implemented by an algebraic data type in Haskell, where \C!Nil! denotes an empty list and \C!Cons h t! is a list that comprises a first element \C!h! and a list of remaining elements \C!t!.
\begin{lstlisting}[style=haskell]
data List $\alpha$ = Nil | Cons $\alpha$ (List $\alpha$)
\end{lstlisting}
Then, the concatenation of two lists can be recursively defined by the function \C!concat! below.
\begin{lstlisting}[style=haskell]
concat :: List #a -> List #a -> List #a
concat Nil l = l
concat (Cons x xs) l = Cons x (concat xs l)
\end{lstlisting}
Although there are no refinement type annotations in this example, \LH/ automatically establishes termination and totality of \C!concat!.
It is straightforward to show that the length of the concatenation of two lists is equal to the sum of the list lengths.
In order to verify this property, we implement a function \C!len! computing the length of lists.
\begin{lstlisting}[style=haskell]
{-@ type #N = {v:#Z | v >= 0} @-}

{-@ measure len @-}
{-@ len :: List #a -> #N @-}
len Nil = 0
len (Cons x xs) = 1 + len 
\end{lstlisting}
The first line introduces the type \C!$\mathbb{N}$! as a \emph{refinement} of Haskell's integer type \C!Int! with the logical predicate \C!v >= 0!.
That is, all values of type \C!$\mathbb{N}$! are ensured to be integers that are non-negative.
This type is used in \C!len! to enforce the property that the length of a list is always a natural number.
The annotation \C!measure len! declares \C!len! a \emph{measure}.
Measures can be used in other refinement annotations by lifting the function definition to the refinement logic \cite{conf/icfp/VazouSJVJ14}.
This is possible for functions that have exactly one parameter such that the function has exactly one definition for each constructor of the parameter.
Moreover, the function can only structurally recurse on its single argument and may only contain functions that are also part of the refinement logic.
The measures defined for a data type are automatically tracked by \LH/.
%
%The annotation \C!reflect len! allows that \C!len! is used in other refinement annotations by lifting the function definition to the refinement logic.
%Functions annotated with \C!reflect! are also referred to as \emph{reflected}.
%This feature is named \emph{refinement reflection} \cite{ref-refl} and is used extensively in this paper for writing expressive refinement types and allowing manual proofs.
%Refinement reflection is only possible for functions that are total, terminating, and only refer to other functions that are available in the refinement logic.

We can now state the following refinement type for list concatenation, which is then successfully verified by \LH/.
\begin{lstlisting}[style=haskell]
{-@ concat :: l1:List #a -> l2:List #a -> {r:List #a | len r = len l1 + len l2} @-}
\end{lstlisting}
%\LH/ also enables efficient reasoning about sets through an encoding using the SMT theory of functional arrays.
Note that the refinement type for the output value \C!concat! can refer to the inputs, which enables the verification of relational properties between input and output of the function.
This property is fully automatically verified by \LH/.
In this case we speak of \emph{intrinsic} verification.

Properties that require manual proofs can be established by \emph{extrinsic} verification.
In such cases, one can write a function that computes proof terms witnessing the validity of the statement.
A proof in \LH/ is represented as the unit type \C!()! refined with the logical predicate encoding the theorem statement, also denoted as \emph{proof term}.
Assume one wants to prove that for all lists \C!l1! and \C!l2! it holds that a value is contained in \C!concat l1 l2! if and only if it is contained in either of the lists.
To this end, it is necessary to lift the definition of \C!concat! into the refinement logic.
Although \C!concat! does not satisfy the requirements of a measure, the function can be lifted using \emph{refinement reflection} \cite{ref-refl}.
This entails to add the annotation \C!{-@ reflect concat @-}! to the function definition.
Refinement reflection is possible for all total and terminating functions that only refer to other functions already available in the refinement logic.

Now that \C!concat! is reflected, a theorem that proves the aforementioned fact can be given below.
\begin{lstlisting}[style=haskell]
{-@ reflect $\in$ @-}
{-@ ($\in$) :: Eq #a => v:#a -> l:List #a -> {r:#B | r => (len l > 0)} @-} ~\label{lst:in}~ 
v $\in$ Nil = False
v $\in$ (Cons (Outcome c x p) xs) = v == x || v $\in$ xs

{-@ thm_concat_in :: Eq #a => v:#a -> l1:List #a -> l2:List #a ->
        { v $\listIn/$ (concat l1 l2) <=> (v $\listIn/$ l1 || v $\listIn/$ l2) } @-}
thm_concat_in v Nil l2 = ()
thm_concat_in v (Cons x xs) l2 = thm_concat_in v xs l2
\end{lstlisting}
The plain Haskell type of \C!thm_concat_in! is \C!List #a -> ()!.
In refinement annotations, \C!{ P }! is syntactic sugar for \C!{ v:() | P }!.
Hence, \C!thm_concat_in v l1 l2! computes a proof term certifying the validity of the logical formula \C!v $\listIn/$ (concat l1 l2) <=> (v $\listIn/$ l1) || (v $\listIn/$ l2)!.

The proof goes by induction on the list \C!l1!.
The base case, i.e., \C!l1 = Nil!, is proved automatically.
If \C!l1 = Cons x xs!, it suffices to appeal to the induction hypothesis via a recursive call to the theorem.
\LH/ makes short proofs like this possible using a proof search algorithm called \emph{Proof By Logical Evaluation} that unfolds and reduces expressions as needed \cite{ref-refl}.

It is also possible to formalise complex mathematical statements.
For instance, during the analysis of the probabilistic algorithms presented in this paper, we need to manipulate finite summations and logarithms, resorting to vastly different approaches.
That is, reasoning about finite sums is performed through refinement reflection and a small library of theorems, while logarithms need to be axiomatised.
First, the finite summation library is illustrated, after which the axiomatic approach for logarithms is presented.

\subsection{Finite Summations}
\label{subsec:finite-summations}
The manual proofs for deriving upper bounds of expected costs presented in sections \ref{sec:rand-quick} and \ref{sec:zip-trees} require manipulation of finite summations, i.e., $\sum_{i=n}^m f(i)$.
We therefore extract the functionality supporting reasoning about finite summations into a separate module, enabling code reuse.
The module consists of a function used to encode a summation together with a collection of useful theorems.
With the goal of capturing the semantics of mathematical summations, we introduce the function \C!finSum! (see \Cref{fig:finSum}).

\begin{figure}
\begin{lstlisting}[style=haskell]
{-@ type Range LO HI = {num:#Z | LO <= num && num <= HI} @-} ~\label{lst:Range}~

{-@ reflect finSum @-}
{-@ finSum :: lo:#Z -> hi:#Z -> f:(Range lo hi -> #R) -> ~\label{lst:finSum}~
        {r:#R | (lo <= hi) => ((r = (f lo) + (finSum (lo + 1) hi f)) &&
                              (r = (finSum lo (hi-1) f) + (f hi)))}
        / [if lo <= hi then hi-lo else 0] @-}
finSum lo hi f | lo == hi = f lo
               |  lo < hi = finSum lo (hi-1) f |> f lo + finSum (lo+1) hi f
               | otherwise = 0

{-@ reflect $\triangleRight$ @-}
{-@ ($\triangleRight$) :: #a -> x:#b -> {r:#b | r = x} @-} ~\label{lst:triangle}~
_ $\triangleRight$ x = x
\end{lstlisting}
\caption{Implementation of Finite Summations through \C!finSum!}
\label{fig:finSum}
\end{figure}

The implementation of \C!finSum! makes use of the operator \C!|>!, which just reduces to its second argument.
The use case of this operator is that in an expression \C!$\Psi$ |> $\Phi$! the logical predicates occurring in the refinement type of $\Psi$ are used to verify $\Phi$.
For a practical example, consider its use in the \C!finSum! function.
The presence of the expression \C!finSum lo (hi-1) f! enables the verification of the fact that \C!finSum lo hi! can also be unfolded from above to \C!(finSum lo (hi-1) f) + (f hi)!.
From now on we write $\finSum{lo}{hi}\ \mathtt{f}$ for \C!finSum lo hi f!. 

The syntax \C!/ [if lo <= hi then hi-lo else 0]! is used to aid \LH/ for proving termination.
Without guidance from the user, termination of this function cannot be shown automatically since \C!lo! increases in one call, while \C!hi! stays the same.
With the syntax \C!/ [$e_1$, $\ldots$, $e_n$]! a \emph{termination metric} can be provided by the user that makes it possible to establish the termination of a function as described next.
Every expression $e_i$ can refer to the function arguments and must always evaluate to a natural number.
Furthermore, for each recursive call, the expressions $(e_1, \ldots, e_n)$ must decrease lexicographically, which proves termination \cite{tp-for-all}.

To see how this works in practice, consider the termination metric of \C!finSum!.
There are two recursive calls, where once \C!lo! stays the same and \C!hi! decreases by one and once \C!lo! increases by one and \C!hi! remains unchanged.
Since both cases happen only if \C!lo < hi! the termination metric evaluates to \C!hi-1-lo! in one case and to \C!hi-(lo+1)! in the other case.
With the help of the SMT solver, it is then inferred that the metric in fact decreases at each call and is always a natural number.

The theorems provided by the finite sum module allow reasoning about summations (see \Cref{fig:finSum-theorems}).
In the statement of \thmSumPermute/ we make use of the function composition operator $\circ$ defined in \LH/ as:
\begin{lstlisting}[style=haskell]
{-@ reflect $\circ$ @-}
{-@ ($\circ$) :: (#b -> #c) -> (#a -> #b) -> (#a -> #c) @-} ~\label{lst:comp}~
($\circ$) f g x = f (g x)
\end{lstlisting}

We present the proof of \thmSumFactor/ here.
The theorem \thmSumFactor/ allows replacing \C!$\HfinSum{lo}{hi}$ f! by \C!c * $\HfinSum{lo}{hi}$ g!, provided that \C!f i = c * g i! for each \C!lo <= i <= hi!.
\begin{lstlisting}[style=haskell]
{-@ thm_sum_factor :: lo:#Z ->  hi:#Z -> c:#R ->
        f:(~\Range/~ lo hi -> #R) -> g:(i:~\Range/~ lo hi -> {r:#R | c * r = f i}) ->
        { $\HfinSum{lo}{hi}$ f = c * $\HfinSum{lo}{hi}$ g } / [if lo <= hi then hi-lo else 0] @-}
thm_sum_factor lo hi c f g | lo > hi = ()
                           | lo == hi = g lo |> ()
                           | lo < hi = g lo |> thm_sum_factor (lo+1) hi c f g
\end{lstlisting}
Its proof first performs a case split.
For \C!lo > hi! both sums reduce to \C!0! and thus a proof is automatically found.
If \C!lo = hi!, it holds that \C!$\HfinSum{lo}{hi}$ f! is equal to \C!f lo!.
By exploiting the refinement annotations of \C!g lo! it is possible to show the desired fact.
In the last case the sums can be automatically unfolded one step and an application of the induction hypothesis completes the proof.

\begin{figure}
\begin{minipage}[t]{0.48\linewidth}%
\begin{lstlisting}[style=haskell,xleftmargin=0pt,aboveskip=0pt,belowskip=0pt]
thm_sum_factor :: lo:#Z -> hi:#Z -> ~\label{lst:thm_sum_factor}~
    c:#R -> f:(~\Range/~ lo hi -> #R) ->
    g:(i:~\Range/~ lo hi -> {v:#R | c * v = f i}) ->
    { $\HfinSum{lo}{hi}$ f = c * $\HfinSum{lo}{hi}$ g }

thm_sum_constant :: lo:#Z -> ~\label{lst:thm_sum_constant}~
    {hi:#Z | lo <= hi} -> c:#R ->
    f:(i:~\Range/~ lo hi -> {v:#R | v = c}) ->
    { $\HfinSum{lo}{hi}$ f = c * (hi - lo + 1) }

thm_sum_rewrite :: lo:#Z -> hi:#Z -> ~\label{lst:thm_sum_rewrite}~
    f:(~\Range/~ lo hi -> #R) ->
    g:(i:~\Range/~ lo hi -> {v:#R | v = f i}) ->
    { $\HfinSum{lo}{hi}$ f = $\HfinSum{lo}{hi}$ g }

thm_sum_reverse :: to:#N -> ~\label{lst:thm_sum_reverse}~
    g:(~\Range/~ 0 to -> #R) -> 
    f:(i:~\Range/~ 0 to -> {v:#R | v = g (to-i)}) ->
    { $\HfinSum{0}{to}$ g = $\HfinSum{0}{to}$ f }
\end{lstlisting}
\end{minipage}%
\hfill\vrule width 0.6pt\hfill
\begin{minipage}[t]{0.48\linewidth}%
\begin{lstlisting}[style=haskell,xleftmargin=0pt,aboveskip=0pt,belowskip=0pt]
thm_sum_linear :: lo:#Z -> hi:#Z -> ~\label{lst:thm_sum_linear}~
    left:(~\Range/~ lo hi -> #R) ->
    right:(~\Range/~ lo hi -> #R) -> 
    f:(i:~\Range/~ lo hi ->
        {v:#R | left i + right i}) ->
    { $\HfinSum{lo}{hi}$ f = $\HfinSum{lo}{hi}$ left + $\HfinSum{lo}{hi}$ right }

thm_sum_shift :: lo:#Z -> hi:#Z -> k:#N -> ~\label{lst:thm_sum_shift}~
    g:(~\Range/~ (lo+k) (hi+k) -> #R) ->
    f:(i:~\Range/~ lo hi -> {v:#R | v = g (i+k)}) ->
    { $\HfinSum{lo}{hi}$ f = $\HfinSum{lo+k}{hi+k}$ g }

thm_sum_permute :: n:#N -> f:(~\Range/~ 0 n -> #R) -> ~\label{lst:thm_sum_permute}~
    $\pi$:(~\Range/~ 0 n -> ~\Range/~ 0 n) ->
    $\tau$:(~\Range/~ 0 n -> ~\Range/~ 0 n) ->
    (x:~\Range/~ 0 n ->
        {$\pi$ ($\tau$ x) = x && $\tau$ ($\pi$ x) = x}) ->
    { $\HfinSum{0}{n}$ f = $\HfinSum{0}{n}$ (f $\circ$ $\pi$) }
\end{lstlisting}
\end{minipage}

    \caption{Finite Sum Theorems}
\label{fig:finSum-theorems}
\end{figure}

\subsection{Axiomatisation of Logarithms}
\label{subsec:axiom-logarithms}

We support reasoning about logarithms with base 2 through an axiomatic approach.
Since \LH/ uses the SMT theory of real numbers to reason about Haskell's \C!Double! values, the real mathematical logarithm cannot be implemented by a terminating function.
Nonetheless, verification of logarithmic properties is possible by introducing the \emph{uninterpreted} function \logTwo/ together with a set of assumptions satisfied by \logTwo/.
An uninterpreted function can be declared on the logic level using the \C!measure! keyword.
\begin{lstlisting}[style=haskell]
{-@ measure log2 :: #Z -> #R @-} ~\label{lst:logTwo}~
\end{lstlisting}
Axioms can be stated through functions that are annotated with the \C!assume! keyword.
Since \C!assume! disables refinement type checking for the respective function, axioms should be used carefully since it is possible to cause an inconsistency when declaring contradictory axioms. 
Therefore, we only state the following minimal set of 3 logarithmic properties from which more complex theorems can be derived.
\begin{lstlisting}[style=haskell]
{-@ assume     log2_base :: {~\logTwo/~ 2 = 1} @-}
{-@ assume  log2_product :: {x:#Z | 0 < x} -> {y:#Z | 0 < y} ->
                            {~\logTwo/~ (x * y) = ~\logTwo/~ x + ~\logTwo/~ y} @-}
{-@ assume log2_monotone :: {x:#Z | 0 < x} -> {y:#Z | x <= y} -> {~\logTwo/~ x <= ~\logTwo/~ y} @-}
\end{lstlisting}
Note that in these assumptions it is ensured that \logTwo/ is only applied to positive arguments.
Therefore \C!$\logTwo/$ 0! stays undefined as desired, since there is no assumption that can be applied in this case.
We now show these assumptions can be used to establish various logarithmic identities that will be put to use in the case studies.

First, we establish that \C!$\logTwo/$ 1 = 0! by using \C!log2_product 2 1!, which results in the fact that \C!$\logTwo/$ 2! is equal to \C!$\logTwo/$ 2 + $\logTwo/$ 1!.
The equality can be transformed into \C!1 = 1 + $\logTwo/$ 1! by using \C!log2_base!, from which \C!$\logTwo/$ 1 = 0! is evident.
\begin{lstlisting}[style=haskell]
{-@ log2_1 :: {~\logTwo/~ 1 = 0} @-}
log2_1 = log2_product 2 1 |> log2_base
\end{lstlisting}

\noindent
This fact is then used in the theorem \C!log2_ge_0!, which verifies by induction that the logarithm of a positive integer is always non-negative.
\begin{lstlisting}[style=haskell]
{-@ log2_ge_0 :: {x:#Z | x > 0} -> {~\logTwo/~ x >= 0} @-} ~\label{lst:log2_ge_0}~
log2_ge_0 x = if x == 1 then log2_1 else log2_ge_0 (x-1) |> log2_monotone x (x-1)
\end{lstlisting}
In the case that \C!x > 1! the induction hypothesis is used together with the monotonicity of \logTwo/.

We end this section with a useful logarithmic inequality that we employ in two of the case studies:
\[ 2 + \log_2(x) + \log_2(y) \leqslant 2\log_2(x+y) \quad \text{ for } x, y \geqslant 1. \]
This inequality has also been used by Okasaki \cite{Okasaki:1999}.
In the formalised proof, presented below, we have to instantiate the needed assumptions one after another such that the SMT solver is able to verify the inequality from the given facts.
For each used axiom we state in the comment how the information is used to gradually build the proof.
\begin{lstlisting}[style=haskell]
{-@ log2_inequality :: {x:#Z | 0 < x} -> {y:#Z | 0 < y} -> ~\label{lst:log2_inequality}~
                       {2 + ~\logTwo/~ x + ~\logTwo/~ y <= 2 * ~\logTwo/~ (x+y)} @-}
log2_inequality x y =
       log2_product (x+y) (x+y)            -- $\color{commentGray}\log_2((x+y)\cdot (x+y)) = 2\log_2(x+y)$
    |> log2_monotone (4*x*y) ((x+y)*(x+y)) -- $\color{commentGray}\log_2(4xy) \leqslant 2\log_2(x+y)$
    |> log2_product (4*x) y                -- $\color{commentGray}\log_2(4x) + \log_2(y) \leqslant 2\log_2(x+y)$
    |> log2_product 4 x                    -- $\color{commentGray}\log_2(4) + \log_2(x) + \log_2(y) \leqslant 2\log_2(x+y)$
    |> log2_product 2 2 |> log2_1          -- $\color{commentGray}2 + \log_2(x) + \log_2(y) \leqslant 2\log_2(x+y)$
\end{lstlisting}
In the call to \C!log2_monotone! the inequality $4xy \leqslant (x+y)^2$ is automatically validated by the SMT solver.
In a pen-and-paper proof (cf. \cite[Lemma 16]{Hofmann22}) one can argue that $(x+y)^2 - 4xy = (x-y)^2 \geqslant 0$.

\section{Probability Monad}
\label{sec:prob-monad}

Probabilistic algorithms, i.e., functions whose output is a probability distribution, are usually implemented using \emph{probability monads} in Haskell (cf. \cite{Ramsey02, Erwig06, Scibior15, VasilenkoVB22}).
This approach enables the succinct representation of randomised programs by structuring computations using a monad.
We implement a probability monad and use \LH/ to provide support for auto-active expected cost and expected value analysis of probabilistic algorithms.
In order to guarantee soundness of the facts established by refinement types and extrinsic theorems, a restriction to \emph{finite} discrete probability distributions is necessary.
In \Cref{sec:inf-dist} we show how to work with infinite distributions and highlight problems arising thereof.
Moreover, in our formalisation of zip trees in \Cref{sec:zip-trees} we introduce a sound way of reasoning about infinite distributions using finite approximations.

\paragraph{Encoding Discrete Probability Distributions}
A discrete probability distribution over type $\alpha$ is represented as a list of outcomes.
An outcome is a tuple $(c, v, p)$, where $c\in\Rnneg$ denotes the cost of the outcome, $v:\alpha$ its value, and $p \in(0,1]$ its probability.
Note that the probability of an outcome is enforced to be positive using refinement types.
\begin{lstlisting}[style=haskell]
{-@
type #Rnneg = {v:#R | v >= 0}
data Outcome #a = Outcome { cost :: #Rnneg
                         , value :: #a
                         , prob  :: {v:#R | 0 < v && v <= 1} }
type Dist #a = List (Outcome #a)
@-}
\end{lstlisting}

\noindent
In a \emph{proper} distribution the probabilities sum up to exactly one.
If the sum is less or equal to one, we speak of a \emph{subdistribution}. 
These notions are formalised by the following types \ProperDist/ and \SubDist/.
The probability monad is designed in such a way that the combination of distributions is proper if all the inputs are also proper distributions.
\begin{lstlisting}[style=haskell]
{-@ reflect sumProb @-}
{-@ sumProb :: Dist #a -> #Rnneg @-} ~\label{lst:sumProb}~
sumProb Nil = 0
sumProb (Cons (Outcome _ _ p) xs) = p + sumProb xs

{-@ type ProperDist #a = {d:Dist #a | sumProb d = 1} @-} ~\label{lst:ProperDist}~
{-@ type SubDist #a = {d:Dist #a | sumProb d <= 1} @-} ~\label{lst:SubDist}~
\end{lstlisting}
Note that \sumProb/ can be used in refinement annotations because of refinement reflection enabled by \C!reflect sumProb!.
All the functions defined in this section are reflected, and we thus omit the \C!reflect! annotations.

\paragraph{Motivation for the Encoding}
Formally, a discrete subdistribution is a function $\mu : A \to [0,1]$, where $A$ is a countable set and $\sum_{a\in A} \mu(a) \leqslant 1$.
The \emph{support} $\mathsf{supp}(\mu)$ of a subdistribution $\mu$ is defined by $\{a\in A \mid \mu(a) > 0\}$
The \emph{expected value} of a random variable $f : A \to \Real$ is given by $\sum_{a\in\mathsf{supp}(\mu)} \mu(a)\cdot f(a)$.
It would be possible to represent discrete distributions in \LH/ by a function \C!#a -> {v:#R | 0 <= v && v <= 1}!.
However, problems arise when trying to model the support of a distribution as well as the expected value.
If distributions were encoded by a function type \C!#a -> {v:#R | 0 <= v && v <= 1}!, this would entail that one has to enumerate all possible values of the abstract type $\alpha$ in order to collect all values with positive probability for computing the support or the expected value.
Thus, the consequence is nontermination and therefore no hope for any form of automation.

Therefore, consider again or formalisation using lists of outcomes.
In order to achieve a high degree of automation, we keep track of expected costs and expected values of distributions throughout the computations carried out by a program.
A simple and \emph{terminating} implementation of these functions is essential such that \LH/ is able to automatically reason about their changes imposed by functions provided by our probability monad.
To this end, the expected cost of a distribution is implemented recursively on the list of outcomes of the distribution.
For expected value calculation, a random variable encoded as a function to $\Real$ is used to transform the value of an outcome into a real number.

\noindent
\begin{minipage}{\linewidth}
\vspace{0.7em}
\begin{minipage}{0.43\textwidth}
\begin{lstlisting}[style=haskell,xleftmargin=0.2cm]
{-@ expectCost :: Dist #a -> #Rnneg @-} ~\label{lst:expectCost}~
expectCost Nil = 0
expectCost (Cons (Outcome c _ p) xs) =
	c * p + expectCost xs
\end{lstlisting}
\end{minipage}
\hfill\vrule width 0.6pt\hfill
\begin{minipage}{0.55\textwidth}
\begin{lstlisting}[style=haskell,xleftmargin=0.5cm]
{-@ expectVal :: (#a -> #R) -> Dist #a -> #R @-} ~\label{lst:expectVal}~
expectVal _ Nil = 0
expectVal f (Cons (Outcome _ x p) xs) =
	(f x) * p + expectVal f xs
\end{lstlisting}
\end{minipage}
\vspace{0.7em}
\end{minipage}

From now on, we write \C!$\Ecost{\mathtt{d}}$! for \C!expectCost d!, and \C!$\Eval{\mathtt{f}}{\mathtt{d}}$! for \C!expectVal f d!.
In the following we provide a high-level introduction to the functions provided by the probability monad.

Additionally, support and the probability of arbitrary values can be neatly represented using the list-based encoding as implemented by the functions \C!support! and \probOf/ below.

\noindent
\begin{minipage}{\linewidth}
\vspace{0.7em}
\begin{minipage}{0.43\textwidth}
\begin{lstlisting}[style=haskell,xleftmargin=0.2cm]
{-@
support :: Dist #a -> List #a
@-}
support Nil = Nil
support (Cons (Outcome c v p) xs) =
    Cons v (support xs)
\end{lstlisting}
\end{minipage}
\hfill\vrule width 0.6pt\hfill
\begin{minipage}{0.55\textwidth}
\begin{lstlisting}[style=haskell,xleftmargin=0.5cm]
{-@ probOf :: Eq #a => x:#a -> d:Dist #a -> ~\label{lst:probOf}~
      {p:#Rnneg | p <= ~\sumProb/~ d &&
               ((x $\listIn/$ support d) <=> (p > 0))} @-}
probOf _ Nil = 0
probOf v (Cons (Outcome _ x p) xs) =
    (if x == v then p else 0) + (probOf v xs)
\end{lstlisting}
\end{minipage}
\vspace{0.7em}
\end{minipage}

A nice property verified in the refinement annotations of \probOf/ is that a value \C!x! is included in the support of a distribution if and only if the probability of \C!x! in the distribution is positive.

\paragraph{Core Functions of the Probability Monad}
We begin with the simplest function that just wraps a value into a singleton distribution. 
\begin{lstlisting}[style=haskell]
{-@ dunit :: f:(#a -> #R) -> x:#a -> ~\label{lst:dunit}~
             {r:ProperDist #a | $\Ecost{\mathtt{r}}$ = 0 && $\Eval{\mathtt{f}}{\mathtt{r}}$ = f x} @-}
dunit _ x = Cons (Outcome 0 x 1) Nil
\end{lstlisting}
Many functions of the probability monad take a random variable as input in order to express how the expected value of it changes with respect to the created distribution.
The expected value of \C!f! with respect to \C!dunit f x! is automatically verified as being equal to \C!f x!.
In programs where we are not interested in expected values of a distribution, we omit the random variable parameter.

Costs are incurred using the \tick/ function.
Passing a random variable \C!f! admits the verification that expected values are unchanged by \tick/.
\begin{lstlisting}[style=haskell]
{-@ tick :: f:(#a -> #R) -> c:#Rnneg -> d:Dist #a -> ~\label{lst:tick}~
            {r:Dist #a | ~\sumProb/~ r = ~\sumProb/~ d && $\HEval{\mathtt{f}}{\mathtt{r}}$ = $\HEval{\mathtt{f}}{\mathtt{d}}$ &&
                        $\HEcost{\mathtt{r}}$ = c * (~\sumProb/~ d) + $\HEcost{\mathtt{d}}$} @-}
tick f c Nil = Nil
tick f c (Cons (Outcome c' x p) xs) = Cons (Outcome (c+c') x p) (tick f xs)
\end{lstlisting}
If \C!d! is a proper distribution, then \C!tick f c d! is also a proper distribution.
And as \C!$\sumProb/$ d = 1! it follows that \C!$\HEcostOpen$tick f c d$]$ = c + $\HEcostOpen$d$]$!. 

Simple combination of distributions is realised by just appending the list of outcomes of one distribution after another list of outcomes.
A useful property of this operation is that the expected costs and expected values of the resulting distribution can be calculated as the sum of the input distributions.
Moreover, also the sum of probabilities of the output distribution is the sum of probabilities of the inputs.
\begin{lstlisting}[style=haskell]
{-@ append :: f:(#a -> #R) -> d1:Dist #a -> d2:Dist #a -> ~\label{lst:append}~
              {r:Dist #a | ~\sumProb/~ r = ~\sumProb/~ d1 + ~\sumProb/~ d2 &&
                          $\HEcost{\mathtt{r}}$ = $\HEcost{\mathtt{d1}}$ + $\HEcost{\mathtt{d2}}$ && $\HEval{\mathtt{f}}{\mathtt{r}}$ = $\Eval{\mathtt{f}}{\mathtt{d1}}$ + $\Eval{\mathtt{f}}{\mathtt{d2}}$} @-}
append Nil d2 = d2
append (Cons x xs) d2 = Cons x (append xs d2)
\end{lstlisting}
Previously, we mentioned that the resulting distributions of functions provided by the monad should be proper provided that all distributions provided as input are as well.
But it now happens that the sum of probabilities of \C!append f d1 d2! would be equal to two if \C!d1! and \C!d2! are proper.
Therefore, there has to be a way to scale the probabilities in order for the sum of probabilities to not exceed one.
The function \pMap/ takes care of this issue.
\begin{lstlisting}[style=haskell]
{-@ pMap :: f:(#a -> #R) -> {p:#R | 0 < p && p <= 1} -> d:Dist #a -> ~\label{lst:pMap}~
            {r:Dist #a | ~\sumProb/~ r = p * ~\sumProb/~ d &&
                        $\HEcost{\mathtt{r}}$ = p * $\HEcost{\mathtt{d}}$ && $\HEval{\mathtt{f}}{\mathtt{r}}$ = p * $\HEval{\mathtt{f}}{\mathtt{d}}$} @-}
pMap f p Nil = Nil
pMap f p (Cons (Outcome c x q) xs) = Cons (Outcome c x (p * q)) (pMap f p xs)
\end{lstlisting}
Again, note how \pMap/ relates the expected cost and expected value of the output distribution to the inputs of the function.

Next, we see how \append/ and \pMap/ can be combined.
Often in a probabilistic program, the next computation step depends on the outcome of a single random trial that succeeds with some probability $p$ and fails with probability $1-p$.
This is also known as a \emph{Bernoulli} trial, and we therefore also name the function modelling this situation \C!bernoulli!.
\begin{lstlisting}[style=haskell]
{-@ bernoulli :: f:(#a -> #R) -> {p:#R | 0 <= p && p <= 1.0} -> ~\label{lst:bernoulli}~
        success:Dist #a -> failure:Dist #a ->
        {r:Dist #a | ~\sumProb/~ r = p * (~\sumProb/~ success) + (1-p) * (~\sumProb/~ failure)
                 && $\HEcost{\mathtt{r}}$ = p * $\HEcost{\mathtt{success}}$ + (1-p) * $\HEcost{\mathtt{failure}}$
                 && $\HEval{\mathtt{f}}{\mathtt{r}}$ = p * $\HEval{\mathtt{f}}{\mathtt{success}}$ + (1-p) * $\HEval{\mathtt{f}}{\mathtt{failure}}$} @-}
bernoulli f 0 success failure = failure
bernoulli f 1 success failure = success
bernoulli f p success failure = ~\append/~ f (~\pMap/~ f p success) (~\pMap/~ f (1-p) failure)
\end{lstlisting}
Note that in the cases where \C!p! is either \C!0! or \C!1! we can just return the appropriate distribution as the outcome is certain.
All of the refinement annotations are automatically discharged by \LH/, which is possible because \append/ and \pMap/ keep track of expected cost, expected value and sum of probabilities.
Consider the distribution defined by \C!bernoulli f p d1 d2!, which is proper if both \C!d1! and \C!d2! are proper.
All outcomes of \C!d1! are scaled by \C!p! using \pMap/ whereas all outcomes of \C!d2! are scaled by \C!1-p!.

Sequential composition of distributions is enabled through the \bind/ function.
\begin{lstlisting}[style=haskell]
{-@ bind :: d:Dist #a -> (#a -> ~\ProperDist/~ #b) -> {r: Dist #b | ~\sumProb/~ r = ~\sumProb/~ d} @-} ~\label{lst:bind}~
bind Nil f = Nil
bind (Cons (Outcome c x p) xs) f = ~\append/~ (~\pMap/~ p (tick c (f x))) (~\bind/~ xs f)
\end{lstlisting}
Intuitively, \C!bind d f! computes a distribution as follows.
If \C!d! is the empty distribution, then \C!bind d f! is empty as well.
Otherwise, \C!d! is of the form \C!Cons (Outcome c x p) xs! and we can evaluate \C!f x!.
Since \C!x! has probability \C!p! and cost \C!c!, we add the cost \C!c! using \tick/ and scale the probabilities of \C!f x! using \C!pMap!.
Finally, this distribution is combined using \append/ with the result from calling \bind/ recursively on the remaining list of outcomes \C!xs!.
It is ensured that \C!$\bind/$ d f! is a proper distribution whenever \C!d! is one as well, and only proper distributions are computed by \C!f!.

Lastly, we provide functions allowing to \emph{lift} unary and binary functions into functions over distributions.
More formally, a unary function of type \C!#a -> #b! can be lifted into a function with type \C!Dist #a -> Dist #b!.
For binary functions, the lifting of \C!#a -> #b -> #c! has the corresponding type \C!Dist #a -> Dist #b -> Dist #c!.

Since lifting of unary functions is implemented by just mapping each value of an outcome, we denote this operation by \fMap/.
Moreover, as for binary functions the lifting can be thought of combining two distributions into one, we call this function \combine/.

\noindent
\begin{minipage}{\linewidth}
\vspace{0.7em}
\begin{minipage}[t]{0.39\linewidth}
\begin{lstlisting}[style=haskell,xleftmargin=3pt,aboveskip=0pt,belowskip=0pt]
{-@
fMap :: f:(#a -> #b) -> d:Dist #a -> ~\label{lst:fMap}~
    {r:Dist #b | ~\sumProb/~ r = ~\sumProb/~ d
             && $\HEcost{\mathtt{r}}$ = $\HEcost{\mathtt{d}}$}
@-}
fMap f Nil = Nil
fMap f (Cons (Outcome c x p) xs) =
    Cons (Outcome c (f x) p)
         (fMap f xs)
\end{lstlisting}
\end{minipage}%
\hfill\vrule width 0.6pt\hfill
\begin{minipage}[t]{0.57\linewidth}%
\begin{lstlisting}[style=haskell,xleftmargin=2pt,aboveskip=0pt,belowskip=0pt]
{-@
combine :: f:(#a -> #b -> #c) -> d1:Dist #a -> d2:Dist #b -> ~\label{lst:combine}~
    {r:Dist #c | ~\sumProb/~ r = (~\sumProb/~ d1) * (~\sumProb/~ d2)
             && $\HEcost{r}$ = (~\sumProb/~ d2) * $\HEcost{d1}$
                        + (~\sumProb/~ d1) * $\HEcost{d2}$}
@-}
combine f Nil d2 = Nil
combine f (Cons (Outcome c x p) xs) d2 =
    ~\append/~ (~\pMap/~ p (~\tick/~ c (~\fMap/~ (f x) d2)))
           (combine f xs d2)
\end{lstlisting}
\end{minipage}
\vspace{0.7em}
\end{minipage}

Because \fMap/ only modifies the values of outcomes, the expected cost remains unchanged.
%Intuitively, \C!combine f d1 d2! computes a cartesian product of the outcomes of \C!d1! and \C!d2! and produces a new distribution by applying \C!f! to each of the combinations.
Intuitively, \C!combine f d1 d2! samples one value from \C!d1! and one \C!d2! and applies \C!f! to these values.
The implementation of \combine/ is more involved.
When \C!d1! is of the form \C!Cons (Outcome c x p) xs!, proper handling of probability and cost is achieved by \pMap/ and \tick/.
Additionally, \C!fmap (f x) d2! achives that the value \C!x! is applied to \C!f! for each value in \C!d2!.
If \C!d1! and \C!d2! be proper distributions, then \C!combine f d1 d2! is also proper and the expected cost is given by {\footnotesize $\HEcost{\mathtt{d1}} + \HEcost{\mathtt{d2}}$}.

\paragraph{Monad Laws}
There is a convention in Haskell that implementations of a typeclass should adhere to certain algebraic laws.
In the case of monads, our implementation has to satisfy the following rules.
\begin{itemize}
    \item Left identity: \C!$\bind/$ ($\dunit/$ x) f = f x!.
    \item Right identity: \C!$\bind/$ d $\dunit/$ = d!.
    \item Associativity: \C!$\bind/$ ($\bind/$ m g) h = $\bind/$ m (\x -> $\bind/$ (g x) h)!
\end{itemize}
We proved these facts within \LH/ and provide the proofs in the appendix (\Cref{sec:prob-monad-thms}).
In addition to these laws, the probability monad exhibits several more laws that we can exploit for the analysis of programs.
For instance, consider the program \C!$\bind/$ ($\bernoulli/$ f p d1 d2) g!, for which expected costs need to be analysed.
Using a law of the probability monad, it can be rewritten into the equivalent form \C!$\bernoulli/$ f p ($\bind/$ d1 g) ($\bind/$ d2 g)!.
In addition to the monad laws, our implementation also adheres to the laws for the Haskell applicative and functor typeclasses.

\section{Meldable Heaps}
\label{sec:meldable-heaps}

\setlength{\abovedisplayskip}{2.5pt}
\setlength{\belowdisplayskip}{2.5pt}

In this section we demonstrate the automation and expressivity of our probability monad by implementing and verifying meldable heaps.

Let us reconsider the implementation of the meld operation for randomised meldable heaps presented in \Cref{lst:meldableheaps}.
In this case study we show the functional correctness of \meld/ in a fully automated way.
Moreover, through providing a hint to the solver by calling the external theorem \logTwoIneq/, the optimal logarithmic upper bound on the expected cost is established by \LH/.

In order to show termination of the \meld/ function, we need to provide a termination metric as the function recurses on either of the two input heaps.
To this end, we define a notion of size of heaps and instruct \LH/ to use the sum of the heaps sizes as termination metric (see line~\ref{lst:meld:metric}).
Additionally, for functional correctness we use \LH/'s multiset axiomatisation.
In both cases we can define a measure. 

\begin{minipage}{\linewidth}
\vspace{0.7em}
\begin{minipage}[t]{0.41\linewidth}
\begin{lstlisting}[style=haskell,xleftmargin=3em,belowskip=0pt,aboveskip=0pt]
{-@ measure size @-}
{-@ size :: Heap k -> ~\label{lst:heapSize}~
            {r:#Z | r >= 1} @-}
size Empty = 1
size (Heap _ l r) =
    size l + size r
\end{lstlisting}    
\end{minipage}%
\hfill\vrule width 0.6pt\hfill
\begin{minipage}[t]{0.57\linewidth}
\begin{lstlisting}[style=haskell,xleftmargin=2em,belowskip=0pt,aboveskip=0pt]
import Language.Haskell.Liquid.Bag
           ( Bag, empty, put, union )

{-@ measure bag @-}
{-@ bag :: Ord #a => Heap #a -> Bag #a @-} ~\label{lst:bag}~
bag Empty = empty
bag (Heap k l r) = put k (union (bag l) (bag r))
\end{lstlisting}
\end{minipage}
\vspace{0.7em}
\end{minipage}

The function \C!size! counts the empty leaf nodes of a heap and always results in a positive integer.
Thus, when recursing on either the left or right child of a heap in \meld/, the sum of the sizes of the input heaps can be proved to decrease by at least one.
The multiset of elements contained in a heap over ordered type $\alpha$ is computed by \C!bag!.
In line~\ref{lst:meld:bag} in \Cref{lst:meldableheaps} \C!bag! is used in the refinement annotation to automatically establish that the multiset of the resulting heap is the union of the elements of the input heaps, where \C!Bag_union! is provided by \LH/ to express multiset union.

The derivation of the upper bound for expected costs uses the axiomatisation of logarithms presented in \Cref{subsec:axiom-logarithms}.
If one of the input heaps is empty, the other heap is returned in a singleton distribution by \dunit/.
In this case, there are no costs but since \logTwo/ is uninterpreted, we have to manually apply the theorem \logTwoGeZero/ to provide the fact that \logTwo/ is non-negative for all positive inputs.
It remains to show the validity of the bound for the scenario where both heaps are non-empty.
Let \C!h1 = Heap k1 l1 r1!, and \C!h2 = Heap k2 l2 r2!.
If \C!k1 <= k2!, the heap \C!h2! can be melded with either \C!l1! or \C!r1!.
Using \bernoulli/ a fair coin toss is simulated, where one time \C!l1! is melded with \C!h2! and the other time \C!r1! is melded with \C!h2!.
The operator \tick/ is employed to incur a cost of one for each recursive call.
Moreover, with \fMap/ the resulting heap is used to either replace \C!l1! or \C!r1!.
The case \C!k1 > k2! is symmetric to the above, where either \C!l2! or \C!r2! is melded with \C!h1!.

Hence, assume that \C!k1 <= k2!.
Due to the refinement annotations tracking expected costs, \LH/ is able to deduce the following expected cost 
\footnotesize
\begin{align*}
\HEcostOpen\meld/\ \mathtt{h1\ h2}] ={}& 0.5\cdot(1 + \HEcostOpen\mathtt{\meld/\ l1\ h2}]) + 0.5\cdot(1 + \HEcostOpen\mathtt{\meld/\ r1\ h2}]) \\
                                   ={}& 1 + 0.5\cdot \HEcostOpen\mathtt{\meld/\ l1\ h2}] + 0.5\cdot \HEcostOpen\mathtt{\meld/\ r1\ h2}].
\end{align*}\normalsize

Using the cost bound obtained from the recursive calls, the following is an upper bound to the above expression:
\footnotesize
\[
\HEcostOpen\meld/\ \mathtt{h1\ h2}] \leqslant 1 + 0.5 \cdot \logTwo/\ (\mathtt{size\ l1}) + 0.5\cdot \logTwo/\ (\mathtt{size\ r1}) + \logTwo/\ (\mathtt{size\ h2}).
\]\normalsize
By applying \C!$\logTwoIneq/$ (size l1) (size r1)! (see line \ref{lst:meld:logIneq}) and the fact that \C!size h1! is equal to \C!size l1 + size r1!, we arrive at the desired upper bound
\footnotesize
\[
\HEcostOpen\meld/\ \mathtt{h1\ h2}] \leqslant \logTwo/\ (\mathtt{size\ h1}) + \logTwo/\ (\mathtt{size\ h2}).
\]\normalsize
\section{Infinite distributions: Coupon Collector's Problem}
\label{sec:inf-dist}

So far, the presented probabilistic algorithms resulted only in finite distributions.
That is, the number of computation steps leading to an outcome is guaranteed to be finite.
In this scenario, termination is \emph{absolute} \cite{mciver2005}.
However, many probabilistic programs do not absolutely terminate, as there may exist computation paths that never terminate.
If the probability of all computation paths of infinite lengths is zero, we speak of \emph{almost-sure termination} \cite{Ferrer15}.
In this section we demonstrate using the \emph{coupon collector's problem} that while it is possible to verify simple almost-surely terminating probabilistic programs, automation support is limited and even small mistakes lead to unsoundness.
The formalisation of zip trees presented in \Cref{sec:zip-trees} proposes a sound approach to circumvent these issues.

Given a positive integer $n$, the coupon collector's problem asks for the expected number of coupons one has to buy in order to collect all $n$ different coupons, where the type of a bought coupon is chosen uniformly at random.
We can represent this problem using the \C!collect! function given below.

\begin{lstlisting}[style=haskell]
{-@ lazy collect @-}
{-@ collect :: n:#N -> {i:#N | i <= n} -> ~\ProperDist/~ () @-}
collect n i | i == 0 = ~\dunit/~ ()
            | i <= n  = ~\tick/~ 1 (~\bernoulli/~ (i / n) (collect n (i-1)) (collect n i))
\end{lstlisting}
The two arguments of \C!collect! are the amount of coupon types \C!n! as well as the count of coupon types \C!i! that remain to be collected.
Then, the coupon collector's problem for \C!n! coupon types is represented by \C!collect n n!.
We model the amount of purchased coupons using the cost facility of the probability monad.
Note that \C!collect! is not absolutely terminating because of the recursive calls in the case \C!i < n!.
Since \LH/ has no notion of probabilistic termination, we need to add the annotation \C!lazy collect! in order to disable termination checking for this function.
Such functions are henceforth called \emph{lazy}.

To see how \C!collect! encodes the coupon collector's problem, consider the two different cases of the function.
If \C!i! is zero, all coupon types are collected.
Otherwise, \C!i! out of \C!n! coupons need to be collected.
So we incur a cost of 1 representing a bought coupon.
It happens with a probability of \C!i / n! to purchase a coupon of a yet uncollected type.
In this case, we recurse and decrement \C!i!.
On the other hand, if we bought a coupon of a type that we already gathered, we need to recurse and leave \C!n! and \C!i! unchanged.

In the following, let $C(n,i)$ denote {\footnotesize $\Ecost{\mathtt{collect\ n\ i}}$}.
Using the refinement annotations of the functions provided by the probability monad, we see that the expected cost of \C!collect n i! satisfy the following recurrence:
\[
C(n,0) = 0 \quad\text{and}\quad C(n,i) = 1 + \tfrac{i}{n} C(n,i-1) + (1-\tfrac{i}{n}) C(n,i) \quad \text{if } i > 0
\]
The latter expression can be simplified to $C(n,i) = \frac{n}{i} + C(n,i-1)$.
From $C(n,1) = n$, $C(n,2) = n + \tfrac{n}{2}$, and $C(n,3) = n + \tfrac{n}{2} + \tfrac{n}{3}$ one can extrapolate the solution $C(n,i) = nH_i$, where $H_i$ denotes the $i$-th harmonic number.
Reusing the function \C!harmonic! used for the analysis of randomised quicksort in \Cref{sec:rand-quick}, we can state the following refinement annotation:
\begin{lstlisting}[style=haskell]
{-@ collect :: n:#N -> {i:#N | i <= n} -> {d:~\ProperDist/~ | $\HEcost{\mathtt{d}}$ = n * harmonic i} @-}
\end{lstlisting}
\LH/ is able to automatically verify this cost annotation.
Hence, we have confirmed that the expected amount of purchased coupons is equal to $nH_n$.

While the verification of \C!collect! was straightforward, this is not always the case.
Disabling the termination checker easily yields unsoundness.
Consider the function \C!bad! defined below.

\begin{minipage}{\linewidth}
\begin{lstlisting}[style=haskell]
{-@ lazy bad @-}
{-@ bad :: {d:Dist () | $\HEcost{\mathtt{bad}}$ = 5} @-}
bad = bernoulli 0.5 bad bad
\end{lstlisting}
\end{minipage}

\LH/ happily verifies this function even though the evaluation of \C!bad! never results in any outcomes.
While this example may seem artificial, forgetting base cases in lazy functions can go unnoticed and the refinement annotations are only verified because they are based on circular reasoning.
what is more, while it is possible to use refinement reflection for lazy functions, \LH/ stops to check refinement annotations and lets verification pass without failure for such functions.
As our goal is to retain all the soundness guarantees of \LH/, we propose a sound way to reason about infinite distributions using finite approximations in the next section.
\section{Zip Trees}
\label{sec:zip-trees}

Zip trees, recently introduced by \citet{TarjanLT21}, are ranked binary search trees in which ranks are generated by sampling from a geometric distribution.
They are isomorphic to the \emph{skip list} randomised data structure \cite{Pugh90}.
In comparison to skip lists, the implementation of zip trees is more concise and especially easy in functional languages. 

The idea behind zip trees is to randomly assign heights to nodes in order to simulate the structure of a \emph{perfect} binary tree, i.e.,
a tree where each node either has exactly two children or is a leaf.
In a perfect binary tree the fraction of nodes with a height of $h$ is $1/2^{h+1}$.
Thus, by sampling ranks from a geometric distribution $\mu_{\mathsf{geo}}$ with $\mu_{\mathsf{geo}}(x) = 1/2^{x+1}$ and imposing a max-heap order on the ranks,
it is possible to obtain binary trees that approximate---in expectation within a constant factor---perfect binary trees.

We proceed by first demonstrating the functional correctness of our \LH/ implementation of zip trees in \Cref{subsec:zip-trees-correct}.
Afterwards, we present the formal expected cost analysis carried out within \LH/ in \Cref{subsec:zip-trees-cost}.
This is---to the best of our knowledge--the first formal verification of the expected runtime performance of zip trees.
In addition, we were able to slightly improve the bounds given in the pen-and-paper proof by \citet{TarjanLT21}.

\subsection{Functional Correctness}
\label{subsec:zip-trees-correct}

\begin{figure}
\begin{minipage}[t]{0.49\linewidth}
\begin{lstlisting}[style=haskell,xleftmargin=12pt,xrightmargin=0pt,numbers=left,numbersep=3pt]
{-@
insert :: Ord #a =>
  rank:#N -> key:#a -> root:Tree #a  ->
  {r:Tree #a
    | rk root >= rk (_left r) &&
      rk root >= rk (_right r) &&
      set r = (Set_sng key) $\cup$ set root} ~\label{lst:insert:correct}~
@-}
insert r k Empty = Node r k Empty Empty
insert r k root@(Node rank key left right)
  | k == key = root
  | k < key =
    let t = insert r k left in
    if rk t < rank then root {_left = t}
    else 
      t {_right = root {_left = _right t}} ~\label{lst:insert:rot}~
  | otherwise =
    let t = insert r k right in
    if rk t <= rank then root {_right = t}
    else
      t {_left = root {_right = _left t}}
\end{lstlisting}
\end{minipage}%
\hfill\vrule width 0.5pt\hfill
\begin{minipage}[t]{0.50\linewidth}
\begin{lstlisting}[style=haskell,xleftmargin=12pt,xrightmargin=0pt,numbers=left,numbersep=3pt,firstnumber=last]
{-@
delete :: Ord #a =>
  key:#a -> t:Tree #a ->
  {r:Tree #a
    | rk r <= rk t &&
      set r = Set_dif (set t) (Set_sng key)} ~\label{lst:delete:correct}~
@-}
delete key Empty = Empty
delete key t@(Node _ k l r)
  | key < k =
       thm_not_set_mem_tree key r
    |> t {_left = delete key l}        
  | key > k =
       thm_not_set_mem_tree key l
    |> t {_right = delete key r}
  | otherwise =
       thm_not_set_mem_tree key l
    |> thm_not_set_mem_tree key r
    |> zip key l r
\end{lstlisting}
\end{minipage}

\caption{Zip Trees: Insertion and Deletion}
\label{fig:zip-tree-insert-delete}
\end{figure}

After stating the formal definition of zip trees, we show how to model the data structure in \LH/ using refinement types.
Finally, insertion and deletion algorithms are presented with \emph{intrinsic}, i.e., fully automated, proofs of their functional correctness.

We define a \emph{ranked binary tree} $t$ such that it is either empty, or is non-empty.
A non-empty ranked binary tree $t$ has a left child $t.\mathit{left}$ and a right child $t.\mathit{right}$,
which are both again ranked binary trees; a natural number $t.\mathit{rank}$; and a key $t.\mathit{key}$ of some ordered type,
denoting the rank and the key of the tree's root node, respectively. 
A binary tree $t$ is a \emph{search tree} if either $t$ is empty, or all keys occurring in $t.\mathit{left}$ are less than $t.\mathit{key}$,
all keys occurring in $t.\mathit{right}$ are greater than $t.\mathit{key}$, and both $t.\mathit{left}$ and $t.\mathit{right}$ are binary search trees.

% GM: spacing
%\begin{minipage}{\linewidth}    
\begin{definition}
A \emph{zip tree} $t$ is a ranked binary search tree, such that
% \begin{itemize}
%     \item if $t$ has a non-empty left child, then $t.\mathit{left}.\mathit{rank} < t.\mathit{rank}$, and $t.\mathit{left}$ is a zip tree; and
%     \item if $t$ has a non-empty right child, then $t.\mathit{right}.\mathit{rank} \leqslant t.\mathit{rank}$, and $t.\mathit{right}$ is a zip tree.
% \end{itemize}
(i) if $t$ has a non-empty left child, then $t.\mathit{left}.\mathit{rank} < t.\mathit{rank}$, and $t.\mathit{left}$ is a zip tree; and
(ii) if $t$ has a non-empty right child, then $t.\mathit{right}.\mathit{rank} \leqslant t.\mathit{rank}$, and $t.\mathit{right}$ is a zip tree.
\end{definition}
%\vspace{0.5em}
%\end{minipage}

The invariants of zip trees can smoothly be represented in \LH/ using refinement types as given below.
The refinement type for the left subtree \C!Tree {v:#a | v < _key}! (respectively \C!Tree {v:#a | _key < v}!) impose a search tree structure by constraining the possible values of keys in the subtrees.
Since an empty tree has no rank, we use the \C!rk! measure that returns \C!-1! for empty trees and $t.\mathit{rank}$ otherwise in order to encode the rank ordering conditions of zip trees.

\noindent
\begin{minipage}{\linewidth}
\begin{minipage}[t]{0.71\linewidth}
\begin{lstlisting}[style=haskell,xleftmargin=0pt]
{-@ data Tree #a =
          Empty
        | Node { _rank :: #N, _key :: #a
               , _left :: {l:Tree {v:#a | v < _key} | rk l < _rank }
               , _right :: {r:Tree {v:#a | _key < v} | rk r <= _rank}
               } @-}
\end{lstlisting}
\end{minipage}
\hfill\vrule width 0.6pt\hfill
\begin{minipage}[t]{0.27\linewidth}
\begin{lstlisting}[style=haskell,xleftmargin=5pt]
{-@ measure rk @-}
{-@ rk :: Tree k -> #Z @-}
rk Empty = -1
rk (Node r _ _ _) = r
\end{lstlisting}
\end{minipage}
\vspace{0.2em}

\end{minipage}

% GM
% Similar to the \C!set! measure for lists defined in \Cref{sec:rand-quick}, we define an analogous \C!set! measure that computes the set of keys contained in a zip tree.
We make use of a \C!set! measure that computes the set of keys contained in a zip tree. This measure
is similar to the \C!set! measure for lists defined in \Cref{sec:rand-quick}.

With these definitions in place, it is possible to write the insertion algorithm and have \LH/ automatically verify its functional correctness (see \Cref{fig:zip-tree-insert-delete}).
%\gm{move the listing forward, so that this becomes easier readable; also try to avoid two figures atop each other}
The function \C!inserts! takes as input a rank, key, and a tree in which the rank and key shall be inserted.
The algorithm performs the insertion recursively as follows.
If the input tree---denoted by \C!root! from now on--.is empty, we can replace it with a leaf node that contains the given rank and key.
The algorithm considers three cases for a non-empty tree.
If the key of \C!root! is equal to the input key, then \C!root! can be returned unmodified.
On the other hand, if the key that needs to be inserted is less than the key of \C!root!, the tree \C!t! is defined to be the result of calling the \C!insert! function on the left subtree.
If the rank of \C!t! is less than the rank of the input tree, the result is \C!root {_left = t}!, i.e., the left subtree of \C!root! is replaced by \C!t!.
Otherwise, the rank of \C!t! is greater or equal to the rank of \C!root!.
To satisfy the rank ordering conditions, \C!root! has to be the right subtree of \C!t! and the left subtree of \C!root! is replaced by the right subtree of \C!t! (see line~\ref{lst:insert:rot} of \Cref{fig:zip-tree-insert-delete}).
The remaining case is symmetric.
The refinement annotation on line~\ref{lst:insert:correct} expresses that the set of keys of \C!insert rank key root! is the union of the singleton set containing \C!key! and the key set of \C!root!,
thus establishing the functional correctness of the insertion algorithm.

The deletion algorithm is given on the right of \Cref{fig:zip-tree-insert-delete}.
The extrinsic theorem \C!thm_not_set_mem_tree! defined below is used to establish the functional correctness given in line~\ref{lst:delete:correct}.

\begin{lstlisting}[style=haskell]
{-@ thm_not_set_mem_tree :: key:#a -> t:Tree {v:#a | k $\neq$ key} -> {not (Set_mem key (set t))} @-}
thm_not_set_mem_tree k Empty = ()
thm_not_set_mem_tree k (Node _ _ l r) = thm_not_set_mem_tree k l |> thm_not_set_mem_tree k r
\end{lstlisting}

This theorem is used to assert that a given \C!key! cannot be in the key set of a tree \C!t! that can only contain keys that are not equal to \C!key!, as given by the refinement type of \C!t! in the theorem's type signature.

The algorithm \C!delete! is given a key and a tree from which the node containing the given key shall be removed.
Since the keys in a zip tree enjoy the search tree property, a node can be efficiently searched for.
If the key is found at the root of a tree \C!t!, the \C!zip! function is used to \emph{zip} the left and right subtrees of \C!t! into a single tree, while preserving the zip tree invariants.
In order to zip trees \C!t1! and \C!t2! it has to be the case that the keys of \C!t1! are smaller than the keys of \C!t2!.
This is achieved by the argument \C!k! in the definition of \C!zip! given below that separates the possible key values of the given trees.
Also note that \LH/ again automatically establishes that the key set of \C!zip k t1 t2! is equal to \C!set t1 $\cup$ set t2!.
\begin{lstlisting}[style=haskell]
{-@ zip :: Ord #a => k:#a -> sm:Tree {v:#a | v < k} -> bg:Tree {v:#a | k < v} ->
    {r:Tree #a | rk r = max (rk sm) (rk bg) && set r = set sm $\cup$ set bg}
    / [size sm + size bg] @-}
zip k Empty Empty = Empty
zip k Empty t = t
zip k t Empty = t
zip k sm bg
    | rk sm < rk bg = bg {_left  = zip k sm (_left bg)}
    | otherwise     = sm {_right = zip k (_right sm) bg}
\end{lstlisting}

An important property of zip trees is the \emph{history independence}.
That is, the structure of a zip tree is only dependent on the keys and ranks of its nodes and not on the sequence of insertion and deletion operations.
We have formalised this result in \LH/ and make use of this property in the expected cost analysis presented in \Cref{subsec:zip-trees-cost}.

%\begin{figure}
%\begin{lstlisting}[style=haskell]
%{-@ zip :: Ord #a => k:#a -> sm:Tree {v:#a | v < k} -> bg:Tree {v:#a | k < v} ->
%    {r:Tree #a | rk r = max (rk sm) (rk bg) && set res = set sm $\cup$ set bg}
%    / [size sm + size bg] @-}
%zip k Empty Empty = Empty
%zip k Empty t = t
%zip k t Empty = t
%zip k sm bg
%    | rk sm < rk bg = bg {_left  = zip k sm (_left bg)}
%    | otherwise     = sm {_right = zip k (_right sm) bg}
%\end{lstlisting} 
%
%\caption{Zip Trees: Zip function}
%\label{fig:zip}
%\end{figure}
\subsection{Expected Cost Analysis}
\label{subsec:zip-trees-cost}

In our formalisation of the expected runtime performance of zip trees we managed to follow most of the pen-and-paper proofs given in \citet{TarjanLT21} quite closely.
Our proofs establish that the expected depth of a node in a zip tree containing $n$ nodes is bounded above by $\sfrac{3}{2}\log_2(n) + \sfrac{3}{2}$, a slight improvement over the bound $\sfrac{3}{2}\log_2(n) + \sfrac{7}{2}$ established in \cite{TarjanLT21}.
Using this result, it follows that the expected time for insertion and deletion are both bounded above by $\sfrac{3}{2}\log_2(n) + \sfrac{3}{2}$, where each recursive call is accounted for with a cost of $1$.

% GM: drop, only keep if enough space
The remainder of this section is structured as follows.
We first introduce our approach for dealing with infinitely supported distributions, as we need to reason about the geometric distribution.
Next, the proof for bounding the expected value of the rank of a zip tree is given.
Finally, the bound for the rank is used to establish the expected depth of an arbitrary node of a zip tree.

\paragraph{Encoding Infinite Distributions.}
Zip trees rely on the geometric distribution with a success probability of $1/2$ for generating ranks.
Formally, the geometric distribution $\mu_{\mathsf{geo}} : \Nat \to [0,1]$ with success probability $1/2$ is defined by $\mu_{\mathsf{geo}}(x) = 1/2^{x+1}$.
Hence, the support of $\mu_{\mathsf{geo}}$ is infinite.
Since reasoning about infinite structures easily leads to unsoundness for the reasons listed in \Cref{sec:inf-dist}, we need to find an approach to reason about programs involving infinite distributions in a sound way.

Our goal is to verify bounds on the expected cost and expected values of programs that use infinite distributions.
We explain here how we achieve this in specifically for the geometric distribution.
However, the general approach can be used for arbitrary distributions.

The high-level idea is to show bounds for expected cost and expected values for any \emph{finite approximation} of the geometric distribution.

\begin{definition}
Given $i \in \mathbb{N}$, let $\mu_{\mathsf{geo}}^{< i}(x)$ be defined by $1/2^{x+1}$ if $x < i$ and $0$ otherwise.
We call $\mu_{\mathsf{geo}}^{< i}$ a finite approximation of $\mu_{\mathsf{geo}}$.
\end{definition}

Note that $\mu_{\mathsf{geo}}^{< i}$ coincides with $\mu_{\mathsf{geo}}$ for all values $x < i$ and $\mathsf{supp}(\mu_{\mathsf{geo}}^{< i}) = \{x\in\mathbb{N} \mid x < i\}$.
The distribution $\mu_{\mathsf{geo}}^{< i}$ is not proper for any $i$, as $\sum_{0 \leqslant x < i} 1/2^{x+1} < 1$.
However, since the support is finite, we can safely represent and reason about finite approximations in \LH/.

As we need way to refer to the infinite distribution $\mu_{\mathsf{geo}}$ on the refinement level, we introduce a measure \C!geo! together with the following axioms.
\begin{lstlisting}[style=haskell]
{-@ measure geo :: Dist #N @-}

{-@ assume geo_expectCost :: { $\HEcost{\mathtt{geo}}$ = 0 } @-}
{-@ assume geo_expectVal :: { $\HEval{\mathtt{toReal}}{\mathtt{geo}}$ = 1 } @-}
{-@ assume geo_proper :: { ~\sumProb/~ geo = 1 } @-}
{-@ assume geo_prob :: k:#N -> { ~\probOf/~ k geo = 1 / ~\pow/~ 2 (k+1) } @-}
\end{lstlisting}

With \C!geoFin! we encode the approximations $\mu_{\mathsf{geo}}^{< i}$.

\begin{lstlisting}[style=haskell]
{-@ geoFin :: i:#N -> {d:Dist #N | ~\sumProb/~ d = 1 - (1 / (~\pow/~ 2 i))) &&
                              && ~\sumProb/~ d <= 1} @-}
geoFin 0 = Nil
geoFin i = ~\pow/~ 2 i |> ~\bernoulli/~ 0.5 (~\dunit/~ 0) (~\fMap/~ (+1) (geoFin (i-1))
\end{lstlisting}

Note that by adding the expression \C!pow 2 i! the sum of probabilities of \C!geoFin i! is automatically verified to be \C!1 - (1 / (pow 2 i))!, which corresponds to the sum of probabilities for $\mu_{\mathsf{geo}}$ of the first $i$ values, i.e., $\sum_{0 \leqslant x < i}1/2^{x+1} = 1-2^{-i}$

Taking inspiration from the Monotone Convergence Theorem, we introduce the following axiom.
\begin{definition}\label{def:approx-axiom}
Let \C!prog :: Dist #N -> Dist #a! be a probabilistic program, and let \C!f :: #a -> #Rnneg! be a random variable.
The \emph{approximation axiom} permits establishing \C!$\HEcostOpen$prog geo$]$ <= b! (respectively, \C!$\HEvalOpen{\mathtt{f}}$prog geo$]$ <= b!) if the following conditions are met:
\begin{enumerate}
    \item \label{cond:bound} It holds that \C!$\HEcostOpen$prog (geoFin i)$]$ <= b! (respectively, \C!$\HEvalOpen{\mathtt{f}}$prog (geoFin i)$]$ <= b!) for all $i\in\mathbb{N}$.
    \item \label{cond:mon} The program \C!p! is monotone, i.e., all outcomes of \C!prog (geoFin i)! are included in\\ \C!prog (geoFin (i+1))! for all $i\in\mathbb{N}$.
    \item \label{cond:proper} If \C!d! is a proper distribution, so is \C!prog d!.
\end{enumerate}
\end{definition}

Conditions \ref{cond:bound} and \ref{cond:mon} imply that the sequence \C!$\HEcostOpen$prog (geoFin 0)$]$!, \C!$\HEcostOpen$prog (geoFin 1)$]$!, $\ldots$ is non-decreasing and bounded above by \C!b!.
In order to achieve the same behaviour for expected values, the random variable is required to only produce non-negative real numbers.
Additionally, condition \ref{cond:proper} requires that \C!prog geo! is a proper distribution.
With the help of this axiom, we are able to give bounds on the expected rank and depth for zip trees.

\paragraph{Bounding the Expected Rank.}
The first step of the expected runtime analysis of zip trees is to provide an upper bound of the expected rank of the root node.
We managed to derive a bound $\log_2(n) + 1$ for trees containing $n$ nodes, where $n$ is positive.
A slight improvement over $\log_2(n) + 3$ from \citet{TarjanLT21}.
To this end, we introduce the function \C!insertList! that builds a zip tree from a given list of keys.
As explained above, the distribution from which ranks are sampled is given as a parameter in order to work with any finite approximation of the geometric distribution.
Our formalisation verifies that \C!$\HEvalOpen{\mathtt{\_rank}}$insertList geo l$]$ <= log2 (len l) + 1! using the approximation axiom, allowing to only work with finite approximations of the geometric distribution and the axiomatisation of logarithms given in \Cref{subsec:axiom-logarithms}.

\noindent
\begin{minipage}{\linewidth}
\vspace{0.7em}
\begin{minipage}[t]{0.52\linewidth}
\begin{lstlisting}[style=haskell,xleftmargin=0pt]
{-@
insertList :: Ord #a => d:Dist #N -> l:NoDupList #a ->
  {r:Dist (Tree #a)
    | ~\sumProb/~ r = ~\pow/~ (~\sumProb/~ d) (len l)
   && ((~\sumProb/~ d = 1) => (~\sumProb/~ r = 1))}
@-}
insertList _ Nil = ~\dunit/~ Empty
insertList d (Cons x xs) =
    ~\combine/~ d (insertList g xs) (insert x)
\end{lstlisting}
\end{minipage}%
\hfill\vrule width 0.6pt\hfill
\begin{minipage}[t]{0.47\linewidth}
\begin{lstlisting}[style=haskell,xleftmargin=8pt]
{-@
maxGeo :: d:Dist #N -> n:#N ->
  {r:Dist #N
    | ~\sumProb/~ r = ~\pow/~ (~\sumProb/~ d) n
   && ((~\sumProb/~ d = 1) => (~\sumProb/~ r = 1))}
@-}
maxGeo d 0 = ~\dunit/~ 0
maxGeo d n = ~\combine/~ d (maxGeo d (n-1)) max
\end{lstlisting}
\end{minipage}
\vspace{0.7em}
\end{minipage}

Similar to \C!insertList! we define the function \C!maxGeo d n! that computes the distribution defined by the maximum of \C!n! samples from the distribution \C!d!.
Note that for both functions \LH/ automatically verifies that the sum of probabilities of the resulting distribution is equal to a power of to the sum of probabilities of the distribution passed as an argument (see \Cref{sec:math-fun} for the definition of \pow/).
Additionally, if \C!d! is proper so are \C!insertList d l! and \C!maxGeo d n! for all lists \C!l! and natural numbers \C!n!.
By an extrinsic theorem we establish that \C!$\HEvalOpen{\mathtt{\_rank}}$insertList d l$]$! equals \C!$\HEvalOpen{\mathtt{toReal}}$maxGeo d (len l)$]$! for all distributions \C!d! and lists \C!l!.
That is, the expected value of \C!maxGeo d n! is equal to the expected rank of a zip tree built from \C!n! different keys.

It now remains to be proven that \C!$\HEvalOpen{\mathtt{toReal}}$maxGeo (geoFin i) n$]$ <= $\logTwo/$ n + 1! for every natural number \C!i! and positive integer \C!n!. 
During our formalisation we wanted to build some intuition for the upper bound.
The finite approximation \C!geoFin i! performs at most \C!i! fair coin flips modelled via the \bernoulli/ function and counts the number of failures before a successful trial.
Moreover, during the evaluation of \C!maxGeo (geoFin i) n!, exactly \C!n! such sequences of coin flips are simulated.
It turned out that by reordering the coin flips carried out, we obtain a different program for which the upper bound \C!$\logTwo/$ n + 1! was easier to verify.

To be more precise, instead of generating \C!n! values of \C!geoFin i! one after another, we generate their values in parallel.
Let \C!k = 0! be a variable keeping track of the maximum of the \C!n! generated values.
Then, for \C!i! iterations the following procedure is performed.
At once \C!n! fair coin flips are performed and the value of \C!n! is updated to be the amount of successful trials.
If \C!n! is still positive, the value of \C!k! is incremented.
At the end of the iterations, \C!k! holds the value of the maximum of \C!n! samples from \C!geoFin i!.

The distribution defined by the amount of successful outcomes from \C!n! Bernoulli trials with success probability \C!p! is given by the \emph{binomial distribution}.
The expected value of the binomial distribution with parameters \C!n! and \C!p! is given by \C!n*p!.
Since in our case \C!p = 1/2!, we can infer that in every iteration of the above procedure the value of \C!n! is expected to become \C!n/2!.
The number of steps where \C!n! is replaced by \C!n/2! until \C!n! reaches 0, as tracked by \C!k!, is exactly {\footnotesize $\lfloor \log_2(\mathtt{n}) \rfloor + 1$}.
Therefore, it holds that {\footnotesize $\mathtt{k} \leqslant \log_2(\mathtt{n}) + 1$}.

We implemented this procedure in a function denoted by \C!maxBinom n i! and established through an extrinsic theorem
with the help of our finite summation library and logarithm axiomatisation presented in \Cref{sec:automated-verification} that
\C!$\HEvalOpen{\mathtt{toReal}}$maxBinom n i$]$ <= $\logTwo/$ n + 1! for every natural number \C!i! and positive integer \C!n!.

\paragraph{Relating Isomorphic Distributions.}
A powerful technique for showing relational properties of distributions is proof by \emph{coupling}.
Nevertheless, for reasoning about couplings one either needs a reasonable program logic (cf.~\cite{Barthe_Hsu_2020}) or a set of axioms that can be used to relate distributions (cf.~\cite{VasilenkoVB22}).
Since we only need to show that the distributions given by \C!maxGeo (geoFin i) n! and \C!maxBinom n i! are equivalent, we can use a lightweight approach and show that the distributions contain the same collection of outcomes.
Their order in the list encoding the distribution, however, is different.
If this is the case, we say that distributions are \emph{isomorphic}, since the order of outcomes in the list does not matter semantically.
We fully formalised this aspect in \LH/.

To this end, we provide a theorem which states that the expected value with respect to a random variable is the same for distributions that only differ in the respective ordering of their outcomes.
We show that \C!maxGeo (geoFin i) n! and \C!maxBinom n i! are isomorphic by storing for each outcome the history of results of the Bernoulli trials leading to this outcome.
With this information we establish the following facts:
% GM: space
\begin{itemize}
\item The lengths of \C!maxGeo (geoFin i) n! and \C!maxBinom n i! are equal.
\item The stored history for every outcome is unique in \C!maxGeo (geoFin i) n! and \C!maxBinom n i!.
\item Let {\footnotesize $\ell$} denote the length of \C!maxBinom n i!. Using the stored histories of outcomes we construct
a permutation {\footnotesize $\pi : \{0,\ldots, \ell-1\} \to \{0,\ldots, \ell-1\}$} such that the probability and value at index {\footnotesize $j$} in \C!maxGeo (geoFin i) n! is equal to the probability and value at index {\footnotesize $\pi(j)$} in \C!maxBinom n i!.
\end{itemize}

%\noindent
%\begin{minipage}{\linewidth}
Having verified that \C!maxGeo (geoFin i) n! and \C!maxBinom n i! are isomorphic, we can use the theorem mentioned above to establish the following theorem:
\begin{lstlisting}[style=haskell,xleftmargin=0pt]
thm_maxGeo_maxBinom_isomorphic :: n:#N -> i:#N -> {$\HEvalOpen{\mathtt{toReal}}$maxGeo (geoFin i) n$]$ = $\HEvalOpen{\mathtt{toReal}}$maxBinom n i$]$}
\end{lstlisting}
%\vspace{0.75\medskipamount}
%\end{minipage}

Finally, by combining the above results and applying the approximation axiom it is verified that \C!$\HEvalOpen{\mathtt{\_rank}}$insertList geo l$]$ <= $\logTwo/$ (len l) + 1!.

\paragraph{Bounding the Expected Depth.}
Using the expected bound on the rank of a zip tree it is possible to obtain a bound for the expected depth of a node.
While our proof deriving the expected bound on the rank is quite different from the proof given in \citet{TarjanLT21}, we follow the remaining structure of the pen-and-paper proof quite closely,  while keeping a clean formalisation.

The proofs use the notion of \emph{high} and \emph{low ancestors}.
A node $y$ is an ancestor of a node $x$ in a zip tree, if $y$ occurs on the search path for $x$ starting from the root.
The low ancestors of a node $x$ are all its ancestors with a smaller key (respectively, greater key for the high ancestors).

Following the arguments in \cite{TarjanLT21}, we establish upper bounds on the expected number of low and high ancestors for any arbitrary node in a zip tree.
As these bounds only depend on the rank of the root node, the result \C!$\HEvalOpen{\mathtt{\_rank}}$insertList geo l$]$ <= $\logTwo/$ (len l) + 1! can be applied.
We explain the proof of the bound for the high ancestors.
The argument regarding low ancestors is similar.

Let $x$ be a node in a zip tree $t$.
A node $y$ is a high ancestor of $x$ if and only if $y.\mathit{rank} > x.\mathit{rank}$ and $y.\mathit{key} > x.\mathit{key}$.
Furthermore, the sequence of the high ancestors of $x$ ordered increasing by key is also increasing by rank.
Because of the history independence property of zip trees, we can think of counting the high ancestors of $x$ by changing the order of insertions.
For a pen-and-paper proof it is straightforward to consider the insertions using a different order.
However, the formalisation of these arguments required several extrinsic theorems in order to be able to rearrange the insertions while preserving already established bounds.

The argument goes as follows.
We insert all nodes with keys bigger than the key of $x$ in increasing order one after another into the tree.
Let $k = x.\mathit{rank}$ keep track of the biggest rank encountered so far.
For each newly inserted node $y$ it is checked if $y.\mathit{rank} > k$.
If this is true, $y$ is a high ancestor of $x$ and we update $k$ to be $y.\mathit{rank}$.

Since ranks are sampled from $\mu_{\mathsf{geo}}$, the probability that a rank is at least $k+1$ is $1/2^{k+1}$, i.e., the probability that the outcomes $k+1$ fair Bernoulli trials are all failures.
Hence, we first can perform a Bernoulli trial with success probability of $1/2^{k+1}$ to check if we generate a bigger rank.
If this is the case, the number of high ancestors can be incremented and $k$ is updated to $k + 1 + g$, where $g$ is drawn from $\mu_{\mathsf{geo}}$ and represents the remainder of the geometric trial.
Otherwise, $k$ remains unchanged, and no high ancestor is generated.
Note that anytime we generate a bigger rank, $k$ increases by two in expectation, as the expected value of $\mu_{\mathsf{geo}}$ is one, while the number of high ancestors increases only by one.

Therefore, the expected value of the high ancestors of node $x$ is equal to the expected value of $k/2$.
Because $k$ keeps track of the biggest rank generated for all nodes with keys bigger than $x$, we can use the previously established bound for the expected root rank.
Hence, if $m$ is the number of nodes with keys bigger than $x$, we have that the expected value of the high ancestors of $x$ is at most $\sfrac{1}{2}\cdot(\log_2(m) + 1)$.
Moreover, since $m \leqslant n$ we can generalise the bound to $\sfrac{1}{2}\cdot(\log_2(n) + 1)$ for an arbitrary node.

Similar, for the expected number of low ancestors the only difference is that ranks are only non-increasing, and we increment the number of low ancestors whenever a new rank is at least $k$.
Thus, with probability $1/2^k$ the value of $k$ is incremented by one in expectation, while the number of low ancestors is also incremented by one.
Therefore, the upper bound of the low ancestors of a node is $\log_2(n) + 1$.

Finally, since the ancestors of a node is the sum of its lower and higher ancestors, we can add the derived bounds to obtain an upper bound of 
$\sfrac{3}{2}\cdot(\log_2(n) + 1)$ for the depth of an arbitrary node in a zip tree consisting of $n$ nodes.

%\GM{ideas}
%\begin{itemize}
%\item follows pen-paper proof, inlined with formal representation
%\item emphasise ``poor man coupling'', improved bounds
%\item comparison coupling, we only do isos 
%\item does not require heavy-handed proof calculi, cf. Avanzini et al.
%\end{itemize}

\section{Correctness of Static Probabilistic Cost Analysis}
\label{sec:soundness}

In this section we state and prove the correctness of cost analysis for probabilistic programs in \LH/.
We follow the set-up of the correctness proof for cost analysis as stated in~\cite{HandleyVH20}.

\subsection{Metatheory of \LH/}
\label{subsec:metatheory_-LH}

% \todo[inline]{FZ: Are $\mathit{dunit},\mathit{tick},\mathit{append},\mathit{pMap},\mathit{bernoulli}, \mathit{bind}$ and $\mathit{Dist} \ \tau \mid \mathit{SubDist} \ \tau \mid \mathit{ProperDist} \ \tau$ all the operations and types we want to have in the syntax? Should we include some? Should we delete some?}

\begin{figure}
\begin{tabular}{rccll}
  Constants & $c$ & $\defsymb$ & $n \in \mathbb{Z} \mid s \in \mathbb{R} \mid \mathit{true},\mathit{false} \in \mathbb{B}$ & \\
  & & | & $+,\ -,\ \ldots \mid \ =,\ <,\ \ldots \mid \land,\ \lor,\ \ldots$ & \\
  & & | & $\mathit{dunit},\mathit{tick},\mathit{append},\mathit{pMap},\mathit{fMap},$ &  \\
  & &   & $\mathit{bind},\mathit{combine}$ & (probabilistic extension)\\
  Values & $v$ & $\defsymb$ & $c \mid \lambda x.e \mid D \ \overline{e}$ &\\
  Expressions & $e$ & $\defsymb$ & $v \mid x \mid e \ e \mid \mathit{let} \ x = e \ \mathit{in} \ e$ & \\
              &     &         |  & $\mathit{case} \ e \ \mathit{of} \{D \ \overline{y} \rightarrow e\}$ & \\
  Refinements & $r$ & $\defsymb$ & $e$ & \\
  Basic types & $B$ & $\defsymb$ & $\mathbb{Z} \mid \mathbb{R} \mid \mathbb{B} \mid \mathit{T}$ & \\
  Types & $\tau$ & $\defsymb$ & $\{x:B \mid r\} \ \mid x:\tau_x \rightarrow \tau$ &\\
  & & | & $\mathit{Dist} \ \tau \mid \mathit{SubDist} \ \tau \mid \mathit{ProperDist} \ \tau$ & (probabilistic extension)\\
  Evaluation Contexts & $\evalContextSymb$ & $\defsymb$ & $\bullet \mid \evalContextSymb \ e \mid c \ \evalContextSymb \mid \mathit{case} \ \evalContextSymb \ \mathit{of} \{D \ \overline{y} \rightarrow e\}$ & \\
\end{tabular}
\caption{Syntax of $\lcName$ (core \LH/ plus our probabilistic extensions) }
\label{fig:meta}
\end{figure}

\begin{figure}
\[
\begin{array}{rcl}
\evalContext{e} & \smallStep & \evalContext{e'} \ \ \ \ \text{if} \ e \smallStep e' \\ 
c \ v & \smallStep & \delta(c,v) \\ 
(\lambda x.e) \ e_x& \smallStep & e[e_x/x] \\
\mathit{let} \ x = e_x \ \mathit{in} \ e & \smallStep & e[(\mathit{fix} \ \lambda x.e_x)/x] \\
\mathit{fix} \ e & \smallStep & e (\mathit{fix} \ e) \\
\mathit{case} \ D_j \ \overline{e} \ \mathit{of} \{D_i \ \overline{y_i} \rightarrow e_i\} & \smallStep & e_i[\overline{e}/ \overline{y_i}]
\end{array}
\]
\caption{Small-step Operational Semantics of $\lcName$ (without our probabilistic extensions)}
\label{fig:small-step-semantics}
\end{figure}
In this subsection we review the metatheory of \LH/, compare~\cite{HandleyVH20}.
In \Cref{fig:meta} we state the syntax and semantics of $\lcName$, a language that models the core of \LH/, where we have marked the parts that concern our probabilistic extensions, which we are going to discuss in the next subsection.
The language $\lcName$ includes constants, abstractions, applications, recursive definitions ($\mathit{let} \ x = e \ \mathit{in} \ e$), case statements, and datatypes.
We require all recursive definitions to be terminating; our logic is agnostic to the mechanism used to enforce structural
termination, so we leave this mechanism abstract;
in our implementation we rely on the termination checker of \LH/, which either is able to automatically prove termination based on structural metrics or which can verify the correctness of a user-provided termination metric.
The \emph{operational semantics} of $\lcName$ is defined in \Cref{fig:small-step-semantics} as a contextual, small-step, call-by-name relation $\smallStep$ whose reflective, transitive closure is denoted by $\smallStep^*$.

\smallskip
\emph{Constants.}
Constants applied to values are reduced in one step using the primitive constant
operation $c \ v \smallStep \delta(c,v)$.
For example, consider $(+)$, the primitive addition operator on integers.
In this instance, $\delta(+,n) = +_n$, where $+_n$ is the function that takes some integer $m$ and returns $n+m$.

\smallskip
\emph{Types.}
The \emph{basic types} in $\lcName$ are integers ($\mathbb{Z}$), reals ($\mathbb{R}$), booleans ($\mathbb{B}$) and type constructors (represented by the symbol $\mathit{T}$).
\emph{Types} are either \emph{refinement types} of the form $\{x:B \mid r\}$ where the basic type $B$, captured by the variable $x$, is refined by the boolean expression $r$; or \emph{dependent function types} of the form $x:\tau_x \rightarrow \tau$, where the input $x$ has the type $\tau_x$ and the result type $\tau$ may refer to the binder $x$.

\smallskip
\emph{Denotations.}
Each type $\tau$ denotes a set of expressions $\denotOf{\tau}$, defined by the dynamic semantics in~\cite{conf/icfp/VazouSJVJ14}.
Let $\remRef{\tau}$ the type obtained by erasing all refinements from $\tau$ and $e:\remRef{\tau}$ be the standard typing relation for the $\lambda$-calculus.
Then, we define the denotation of types as follows:
\[
\begin{array}{lcl}
  \denotOf{\{x:B \mid r\}} & \doteq & \{e \mid e:B, \text{ if } e \smallStep^* v, \text{ then } \mid r[v/x] \smallStep^*  \mathit{true} \} \\
  \denotOf{x:\tau_x \rightarrow \tau} & \doteq & \{e \mid e:\remRef{\tau_x \rightarrow \tau}, \text{ forall } e_x \in \denotOf{\tau_x} \text{ it holds that } e \ e_x \in \denotOf{\tau[e_x/x]} \}
\end{array}
\]

\smallskip
\emph{Syntactic Typing}.
The typing judgement $\Gamma \vdash e:\tau$ decides syntactically if e is a member of $\tau$'s denotation using the environment $\Gamma$ that maps variables to their types:
$\Gamma \doteq x_1:\tau_1, \ldots, x_n:\tau_n$.

\smallskip
\emph{Typing of Constants}.
To type a $\lcName$ constant $c$, we use the meta-function $\mathit{Ty}(c)$ that returns the type of $c$, which is used in the following typing axiom:
\[ \overline{\Gamma \vdash c:\mathit{Ty}(c)} \]

To ensure soundness, $\mathit{Ty}(c)$ should satisfy denotational inclusion:
$c \in \mathit{Ty}(c)$.
For example:
\[\begin{array}{lcl}
  \mathit{Ty}(3) & \doteq & \{x:\mathbb{Z} \mid x == 3 \} \\
  \mathit{Ty}(+) & \doteq & x:\mathbb{Z} \rightarrow y:\mathbb{Z} \rightarrow \{z:\mathbb{Z} \mid z == x+y \}
\end{array}
\]
\begin{theorem}[Soundness of Core \LH/~\cite{conf/icfp/VazouSJVJ14}]
\normalfont
If for all constants $c$, $c \in \mathit{Ty}(c)$, then $\emptyset \vdash e::\tau$ implies $e \in \denotOf{\tau}$.
\end{theorem}

\subsection{Correctness of Probabilistic Cost Analysis}

We now state the full definition of $\lcName$, i.e., we give definitions to our probabilistic extensions.

\smallskip
\emph{Extension in terms of core \LH/.}
We define $\mathit{Dist} \ \tau, \mathit{SubDist} \ \tau, \mathit{ProperDist} \ \tau$ as well as $\mathit{dunit}, \mathit{tick}, \mathit{append}, \mathit{pMap}, \mathit{fMap}, \mathit{bind}$, and $\mathit{combine}$ as in \Cref{sec:prob-monad}.
Moreover, the functions $\mathit{expectCost}$, $\mathit{expectVal}$, $\mathit{probOf}$, and $\mathit{support}$ are provided for use in refinement type annotations.
Note that this means that we define our extension in terms of core \LH/.
For example, we define $\mathit{Dist}$ using the datatypes $\mathit{List}$ and $\mathit{Outcome}$ and define $\mathit{SubDist}$ and $\mathit{ProperDist}$ as refinement types of $\mathit{Dist}$.
Note that while we have introduced $\mathit{Dist}$, $\mathit{SubDist}$ and $\mathit{ProperDist}$ in terms of lists of outcomes, we treat them as abstract types in $\lcName$, e.g., we do not allow pattern matching on these types;
rather, the only way to manipulate these types is via the operations $\mathit{dunit}, \mathit{tick}, \mathit{append}, \mathit{pMap}, \mathit{fMap}, \mathit{bind}$, and $\mathit{combine}$.

\smallskip
\emph{Semantics and Soundness of Constants.}
As we define $\mathit{dunit}, \mathit{tick}, \mathit{append}, \mathit{pMap}, \mathit{fMap}, \mathit{bind}$, and $\mathit{combine}$ as core \LH/ terms,
their semantics is induced by the semantics for core \LH/ terms, which we stated in the previous subsection.
Further, we can rely on the (refinement) type checker of \LH/ to check the (refinement) types of our definitions,
i.e., we have $c \in \mathit{Ty}(c)$ for each of the constants $\mathit{dunit}, \mathit{tick}, \mathit{append}, \mathit{pMap}, \mathit{fMap}, \mathit{bind}$, and $\mathit{combine}$.
Therefore, these constants can be \emph{safely} used in $\lcName$ while preserving soundness.

We now obtain the soundness of our probabilistic cost monad as a corollary to the soundness of core \LH/:

\begin{theorem}[Soundness of Probabilistic Cost Analysis]
\normalfont
Let $p: \mathbb{R} \rightarrow \mathbb{B}$ and $q: \mathbb{R} \rightarrow \mathbb{B}$ be predicates, let $f:\tau\rightarrow \mathbb{R}$ be a random variable, and let $e: \{d : \mathit{ProperDist} \ \tau \mid p \ (\mathit{expectCost} \ d) \land q \ (\mathit{expectVal} \ f \ d) \}$ be an expression with $e\smallStep^* v$.

Then, $p \ (\mathit{expectCost} \ v) \smallStep^* \mathit{true}$ and $q \ (\mathit{expectVal} \ f \ v) \smallStep^* \mathit{true}$, as well as $w \in \denotOf{\tau}$ for all $w$ contained in $\mathit{support}\ v$.
\end{theorem}
\section{Related Work}
\label{sec:related-work}

Essentially starting with the seminal work of Kozen~\cite{Kozen81,Kozen:JCSC:85}, there is a large body of work on
the analysis of \emph{probabilistic} programs (see also~\cite{BKS:2020} for further pointers).
Similarly, the literature on the verification of \emph{functional} programs
is extensive (see for example~\cite{EberlHN20} and the references therein). Furthermore, there is
a significant number of works on the (expected) cost analysis of (probabilistic) functional programs,
for example~\cite{NipkowB19,HandleyVH20,WangKH20,LMZ:2021,LMZ:2022}. For brevity, we thus restrict to
closely related work.

\paragraph{Relational properties and coupling}

Typically, probabilistic reasoning is formalised in functional languages
like Haskell through \emph{probability monads}~\cite{Ramsey02, Erwig06, Scibior15, TassarottiH19}.
Similarly, relational properties --- in particular coupling --- can be encoded via a monadic
implementation, cf.~\citet{VasilenkoVB22}. Technical this is obtained in a similar
fashion as our contribution by enhancing \LH/ through a dedicated probability method. In difference
to our work, their monad is wrapping an existing Haskell library. The here established probability monad
is built from first principle. This, in particular, allows us to smoothly express an
expected cost and value analysis, beyond the expressivity of the methodology of~\citet{VasilenkoVB22}. 
A more general formal treatment of relational properties of probabilistic programs is
given by~\citet{AvanziniBDG25}. The central contribution is the development of a dedicated
Hoare logic to this avail. 

\paragraph{Expected cost analysis}

The (automated) cost analysis of probabilistic programs has been intensively studied
(see, for example~\cite{KKMO:ACM:18,NgoCH18,NipkowB19,AMS20,EberlHN20,HandleyVH20,KKM20,WangKH20,LMZ:2021,LMZ:2022,AvanziniMS23}).
In the following, we discuss in more details the literature on cost analysis of probabilistic
functional programs.
\citet{EberlHN20} consider the formalisation of randomised programs, like
\emph{randomised quicksort} and \emph{randomised meldable heaps} in Isabelle.
This clearly relates to our first two case studies.
We note, however, that the improved automated support established by our 
probability monads allows us to derive these formalisations with less overhead. Further, we
only use elementary mathematics.
\citet{WangKH20,LMZ:2022} establish automated support for the analysis of probabilistic (functional) programs.
\citet{LMZ:2022} provides a methodology to synthesise among others an upper bound on the expected cost of \emph{randomised meldable heaps} fully automatically.
In contrast, our treatment of this data structure automatically verifies the
corresponding cost bounds. On the other hand, clearly, our extension of a refinement type system is much more
generally applicable for verification. In particular, the approach of \citet{LMZ:2022} cannot handle the
remaining three case studies.

\paragraph{Randomised algorithms and data structures}

Apart from \emph{randomised quicksort} and \emph{randomised meldable heaps},
we have considered the \emph{coupon collector} scheme and \emph{zip trees}, as case studies.
The coupon collector's problem has also been considered by \citet{AMS20}, where its expected
cost could be fully automatically derived for a variant of the encoding employed here. Note, however, that
this variant of the problem analysed employs a clever abstraction of the algorithm given here. Deriving
the correctness of the abstraction automatically has not been attempted. Similar to zip trees,
\emph{treaps}~\cite{AragonS89}
are binary search trees, whose nodes have a real-valued rank and the nodes are ordered by rank. In contrast
to zip trees, the ranks are chosen uniformly from a continuous distribution.
Treaps are considered in \citet{EberlHN20}, establishing a formal proof of the expected shape of these trees.
\emph{Skip lists}, introduced by~\citet{Pugh90}, are a randomised data structure over lists isomorphic
to zip trees. Skip lists feature a similar (but worse) logarithmic bound on deletion and insertion like
zip trees.

\paragraph{Verification of probabilistic data structures}

\citet{TassarottiH19} have provided a formalisation of two-level concurrent skip lists in Coq.
An expected cost analysis of skip lists has very recently established in~\cite{AvanziniBGMV24}.
\citet{AvanziniBGMV24} uses a weakest pre-expectation calculus in conjunction with a relational Hoare logic~\cite{AvanziniBDG25}
to derive the precise cost bound for deletion and insertion of skip lists. The formalisation is quite
involved and makes use of several abstraction steps, nicely connected to each other through the
use of the relational Hoare logic. Perhaps unavoidably, due to the imperative flavour of skip lists, the
obtained formalisation sets itself apart from the original pen-and-paper proof in~\cite{Pugh90} (see also
related textbook proofs in~\cite{MotwaniRaghavan}).

\section{Conclusion}
\label{Conclusion}

We presented a probability monad for \LH/ that enables the automated verification of expected values and expected costs in probabilistic programs.
Our approach supports discrete distributions with finite support and leverages refinement types to encode probabilistic specifications that are discharged automatically by SMT solvers.
Through four case studies --- including the first formal verification of the expected runtime of zip trees --- we demonstrated both the automation and expressiveness of our framework.

In future work, we aim to extend our monad to support enumerable discrete distributions with infinite support. The current restriction to finite support arises from \LH/’s requirement that functions inside refinement types are terminating.
However, it is sound in principle to allow functions that terminate almost surely;
realizing this would require extending \LH/’s theory and termination checker to reason about probabilistic termination with probability one. 

\bibliographystyle{ACM-Reference-Format}
\bibliography{bib}

\appendix

\section{Appendix}
\label{sec:appendix}

\subsection{Mathematical Functions}
\label{sec:math-fun}

\begin{center}
\begin{minipage}{\linewidth}
\begin{lstlisting}[style=haskell]
{-@
reflect pow
pow :: b:#R -> e:#N -> ~\label{lst:pow}~
    {r:#R | ((e == 0) => (r == 1.0))   &&
            ((b != 0.0) => (r != 0.0)) &&
            ((b > 0.0) => (r > 0.0))   &&
            ((b = 1.0) => (r = 1.0))   &&
            ((b >= 1.0) => (r >= 1.0)) &&
            ((b > 1.0 && e > 0) => (r > 1.0))}
    / [e]
@-}
pow b e
  | e == 0 = 1.0
  | b == 0.0 = 0.0
  | otherwise = b * pow b (e-1)
\end{lstlisting}
\end{minipage}
\end{center}

\begin{center}
\begin{minipage}{\linewidth}
\begin{lstlisting}[style=haskell]
{-@ thm_pow_add :: b:#R -> e1:#N -> e2:#N -> {(pow b (e1+e2)) = ((pow b e1) * (pow b e2))} / [e1] @-}
thm_pow_add b 0 e1 = ()
thm_pow_add b e1 e2 = thm_pow_add b (e1-1) e2
\end{lstlisting}
\end{minipage}
\end{center}

\begin{center}
\begin{minipage}{\linewidth}
\begin{lstlisting}[style=haskell]
{-@ thm_pow_inv :: {b:#R | b != 0.0} -> e:#N ->
        {(1.0 / (pow b e)) = (pow (1.0 / b) e)}
        / [e] @-}
thm_pow_inv b e
    | e == 0 = ()
    | otherwise =
        (1.0 / (pow b e)) |> thm_pow_inv b (e-1)
\end{lstlisting}
\end{minipage}
\end{center}

\subsection{Theorems for Probability Monad Laws}
\label{sec:prob-monad-thms}

\noindent\textbf{Left Identity}

The proof for left identity manually unfolds \C!$\bind/$ ($\dunit/$ x) f! to \C!$\append/$ ($\tick/$ 0 ($\pMap/$ 1 (f x))) ($\bind/$ Nil f)!.
The refinement types for \pMap/ satisfy that \C!$\pMap/$ 1 d! equals \C!d! for every distribution \C!d!.
Similarly, \C!$\tick/$ 0 d! equals \C!d!.
As \C!$\bind/$ Nil f! reduces to \C!Nil!, the aforementioned expression reduces to \C!$\append/$ (f x) Nil!.
Finally, as \append/ is defined in terms of list concatenation, we can employ the theorem \C!thm_listConcat_nil! for \C!f x!, which establishes that the concatenation of a list \C!l! with \C!Nil! is equal to \C!l!.

\begin{lstlisting}[style=haskell]
{-@
thm_left_identity :: x:#a -> f:(#a -> Dist #b) -> { ~\bind/~ (~\dunit/~ x) f = f x }
@-}
thm_left_identity x f = 
       ~\append/~ (~\tick/~ 0 (~\pMap/~ 1 (f x))) (~\bind/~ Nil f)
    |> thm_listConcat_nil (f x)
\end{lstlisting}

\medskip
\noindent\textbf{Right Identity}

The proof for right identity goes by induction, where in the induction step the hypothesis is invoked by calling the theorem recursively with the rest of the distribution.
Both the base case and the induction step are discharged fully automatically by the SMT solver.

\begin{lstlisting}[style=haskell]
{-@
thm_right_identity :: d:Dist #a -> { ~\bind/~ d ~\dunit/~ = d }
@-}
thm_right_identity Nil = ()
thm_right_identity (Cons _ xs) = thm_right_identity xs
\end{lstlisting}

\medskip
\noindent\textbf{Associativity}

The following proof establishing associativity of the probability monad makes use of the \emph{proof combinators} provided by \LH/.
First, the infix function \C!(===)! is used for equational reasoning.
Given \C!$e_1$, $e_2$ : #a!, the expression \C!$e_1$ === $e_2$! reduces to \C!$e_1$!, where the equality \C!$e_1$ = $e_2$! is checked by the SMT solver.
More precisely, the refinement type of \C!$e_1$ === $e_2$! is \C!{v:#a | v = $e_1$ && v = $e_2$}!.
This operator can also be chained, i.e., \C!$e_1$ === $e_2$ === $\cdots$ === $e_n$! satisfies the refinement \C!$e_1$ = $e_n$!, if all intermediate equalities are successfully validated by the SMT solver.

Second, the function \C!(?)! allows referring to other theorems to justify reasoning steps.
For instance, if the extrinsic theorem \C!thm $e_1$! is needed to show the equality between \C!$e_1$! and \C!$e_2$!, one can write \C!$e_1$ ? thm $e_1$ === $e_2$! to provide the needed additional evidence to the solver.

Finally, \C!$e$ *** QED! takes any expression \C!$e$! and always results in the unit value \C!()!.
For theorem proving, the refinement type for expression \C!$e$! needs to be strong enough to show the statement of the current theorem.

We explain the usage of these functions in the proof of the following theorem.

\begin{center}
\begin{minipage}{\linewidth}
\begin{lstlisting}[style=haskell, numbers=left]
{-@
thm_associativity :: d:Dist #a -> f:(#a -> Dist #b) -> g:(#b -> Dist #c) ->
    { ~\bind/~ (~\bind/~ d f) g = ~\bind/~ d (\x:#a -> ~\bind/~ (f x) g) }
@-}
thm_associativity Nil f g = ()
thm_associativity d@(Cons (Outcome c y p) ys) f g = 
        ~\bind/~ d (\x -> ~\bind/~ (f x) g)
    === ~\append/~ (~\pMap/~ p (~\tick/~ c (~\bind/~ (f y) g))) (~\bind/~ ys (\x -> ~\bind/~ (f x) g))
        ? thm_associativity ys f g ~\label{lst:assoc:rec}~
    === ~\append/~ (~\pMap/~ p (~\tick/~ c (~\bind/~ (f y) g))) (~\bind/~ (~\bind/~ ys f) g)
        ? thm_bind_tick c (f y) g ~\label{lst:assoc:bind_tick}~
    === ~\append/~ (~\pMap/~ p (~\bind/~ (~\tick/~ c (f y)) g)) (~\bind/~ (~\bind/~ ys f) g)
        ? thm_bind_pMap p (~\tick/~ c (f y)) g
    === ~\append/~ (~\bind/~ (~\pMap/~ p (~\tick/~ c (f y))) g) (~\bind/~ (~\bind/~ ys f) g)
        ? thm_bind_append (~\pMap/~ p (~\tick/~ c (f y))) (~\bind/~ ys f) g 
    === ~\bind/~ (~\append/~ (~\pMap/~ p (~\tick/~ c (f y))) (~\bind/~ ys f)) g ~\label{lst:assoc:final1}~
    === ~\bind/~ (~\bind/~ d f) g ~\label{lst:assoc:final2}~
    *** QED
\end{lstlisting}
\end{minipage}
\end{center}

The base case is proved automatically.
In the induction step, we manually unfold \C[breaklines]!$\bind/$ d (\x -> $\bind/$ (f x) g)! one time.
In line~\ref{lst:assoc:rec} we use the induction hypothesis to rewrite \C!$\bind/$ ys (\x -> $\bind/$ (f x) g)! to \C!$\bind/$ ($\bind/$ ys f) g!.
Line~\ref{lst:assoc:bind_tick} uses the theorem \C!thm_bind_tick c d f!, which shows that \C!$\tick/$ c ($\bind/$ d f)! is equal to \C!$\bind/$ ($\tick/$ c d) f!.
Similarly, \C!thm_bind_pMap! is used to perform the same step for \pMap/ instead of \tick/.
Then, \C!thm_bind_append d1 d2 f! witnesses that \C!$\append/$ ($\bind/$ d1 f) ($\bind/$ d2 f)! is equal to \C!$\bind/$ ($\append/$ d1 d2) f!.
The final step from line~\ref{lst:assoc:final1} to \ref{lst:assoc:final2} applies to the definition of \bind/ to finish the proof.

\subsection{Randomised Quicksort}
\label{sec:rand-quick}

Quicksort is a divide-and-conquer sorting algorithm that
partitions the input list into smaller parts, depending on a chosen pivot element, and recursively sorts the resulting parts.
While quicksort is fast for random input lists, the order of comparisons degrades to $\bO(n^2)$ in the worst case, if the choices of pivot elements are bad.
Randomised quicksort solves this problem by always choosing pivot elements uniformly at random from the input list, ensuring that the expected number of comparisons is in $\bO(n\log n)$ for any input.

In our formalisation we assume that all elements in the input list are unique.
We implemented randomised quicksort in terms of the probability monad and were able to derive the functional correctness directly using refinement type annotations.
Moreover, we show that the expected number of comparisons for a list of length $n$ is equal to $2(n+1)H_n - 4n$ by a manual proof in \LH/, where $H_n = \sum_{i=0}^{n}1/i$ is the $n$-th harmonic number.
First, we illustrate the functional correctness before we derive the expected cost.

Our implementation makes heavy use of \emph{abstract refinement types} \cite{abs-ref-types}.
\LH/ adds abstract refinement types to Haskell's built-in list data type.
The syntax \C!type T = [#a]<{\x y -> P x y}>! is used to refine a list over elements of type $\alpha$ with a predicate \C!P :: #a -> #a -> Bool!.
Intuitively, if {\footnotesize $[x_1, \ldots, x_n]$} is a list of type \C![#a]<{\x y -> P x y}>!, it is ensured that \C!P $x_i$ $x_j$! is true for all {\footnotesize $i,j$} such that {\footnotesize $1 \leqslant i < j \leqslant n$}.

Using this refined definition of lists, one can define lists that contain no duplicate values as well as lists where consecutive values are increasing:
\begin{lstlisting}[style=haskell]
type NoDupList #a = [#a]<{\x y -> (x != y)}>
type IncrList #a = [#a]<{\x y -> (x < y)}>
\end{lstlisting}
Hence, using \C!IncrList! we can express that the lists in the result distribution are all sorted in increasing order.
However, in addition it needs to hold that each list is a permutation of the input list, i.e., the lists exhibit the same set of elements.
In a similar fashion to the multiset axiomatisation used in \Cref{sec:meldable-heaps}, we can use \LH/'s support for reasoning about sets by defining a measure that can be used in refinement annotations.

\begin{minipage}{\linewidth}
\begin{lstlisting}[style=haskell]
import Data.Set ( Set, empty, union, singleton )

{-@ measure set @-}
{-@ set :: Ord #a => [#a] -> Set #a @-}
set [] = empty
set [x:xs] = union (singleton x) (set xs)
\end{lstlisting}
\end{minipage}

In order to draw numbers from a uniform distribution we introduce the following function.
\begin{lstlisting}[style=haskell]
{-@ uniformBind :: lo:#N -> {hi:#N | lo <= hi} -> ~\label{lst:uniformBind}~
        f:({i:#N | lo <= i && i <= hi} -> ProperDist #a) ->
        {d:ProperDist #a | $\HEcost{\mathtt{d}}$ = (1 / (hi-lo+1)) * $\HfinSum{lo}{hi}$ ($\EcostOnly\ \circ$ f)}
        / [hi-lo] @-}
uniformBind lo hi f =
    | lo == hi = f lo
    | otherwise = ~\bernoulli/~ (1 / (hi-lo+1))
                      (f lo) (uniformBind (lo+1) hi f)
\end{lstlisting}
The intuitive meaning of \C!uniformBind lo hi f! is that a number is chosen uniformly at random between \C!lo! and \C!hi! inclusive, which is then passed to the continuation \C!f!.
Note that the refinement annotation for the result distribution automatically expresses how the expected cost behaves.
This property is automatically verified by \LH/ and will be important in deriving the expected cost of randomised quicksort.

The function \rquick/ in \Cref{fig:rquick} implements the randomised quicksort algorithm.
Its plain Haskell type is \C!Ord #a => [#a] -> Dist [#a]!, i.e., for lists over ordered type $\alpha$, \rquick/ results in a distribution of lists over $\alpha$.
With the help of abstract refinements and the annotations of the probability monad, we can give \rquick/ a more precise type.
That is, given that the input list contains no duplicate values, \LH/ is able to verify that the resulting distribution is proper and all lists in the distribution are sorted in increasing order.
Next, we detail how the functional correctness of the algorithms is verified by \LH/, after which the proof deriving the exact number of expected comparisons is presented.

\begin{figure}[t]
\begin{minipage}[t]{0.49\linewidth}
\begin{lstlisting}[style=haskell,xleftmargin=12pt,xrightmargin=0pt,numbers=left,numbersep=3pt]
{-@
rquick :: Ord #a => ~\label{lst:rquick}~
  l:NoDupList #a ->
  ~\ProperDist/~ ({r:IncrList #a
                | set r = set l
              }) / [len l, 1]
@-}
rquick [] = ~\dunit/~ []
rquick l =
    ~\uniformBind/~ 0 (len l - 1) (rquick_body l)

{-@
rquick_body :: Ord #a => l:NoDupList #a -> ~\label{lst:rquick_body}~
  {i:#N | i < len l} ->
  ~\ProperDist/~ ({r:IncrList #a
                | set r = set l
              }) / [len l, 0]
@-}
rquick_body l i =
  case partition (idx l i) (remIdx l i) of
    (sm, bg) ->
      ~\tick/~ (len l - 1)
           (~\combine/~ (rquick sm) (rquick bg)
                (merge (idx l i)))
\end{lstlisting}
\end{minipage}%
\hfill\vrule width 0.5pt\hfill
\begin{minipage}[t]{0.50\linewidth}
\begin{lstlisting}[style=haskell,xleftmargin=12pt,xrightmargin=0pt,numbers=left,numbersep=3pt,firstnumber=last]
{-@
partition :: Ord #a => x:#a -> ~\label{lst:partition}~
  l:NoDupList {v:#a | v $\neq$ x} -> 
  {t:(NoDupList {r:#a | r < x},
      NoDupList {r:#a | x < r})
    | len l = len (fst t) + len (snd t) &&
      set l = set (fst t) $\cup$ set (snd t)}
@-}
partition _ [] = ([], [])
partition x (y:ys)
  | y > x =
    let (sm,bg) = partition x ys in (sm, y:bg)
  | otherwise =
    let (sm,bg) = partition x ys in (y:sm, bg)

{-@
merge :: Ord #a => x:#a -> ~\label{lst:merge}~
  l1:IncrList {v:#a | v < x} ->
  l2:IncrList {v:#a | x < v} ->
  {l:IncrList #a
    | set l = (Set_sng x) $\cup$ set l1 $\cup$ set l2}
@-}
merge x [] r = x : r
merge x (y:ys) r = y : (merge x ys r)
\end{lstlisting}
\end{minipage}

\caption{Randomised Quicksort}
\label{fig:rquick}
\end{figure}

\paragraph{Functional correctness}
The correctness of the refinement type of \rquick/ can only be established if the actual definition satisfies the type.
To this end, given list \C!l!, the algorithm considers two cases.
First, if \C!l! is empty, the empty list is returned with probability 1 by \C!dunit []! and it is easy to see why the refinement type holds in this case.
Otherwise, \C!l! is non-empty.
Through the \FN{uniformBind} function introduced in \Cref{sec:prob-monad}, a uniformly random natural number \C!i! is chosen between 0 and \C!len l - 1! inclusive and this number is passed to \rquickBody/~\C!l!.
The list element at index $i$ is used as the pivot element.

As \rquick/ and \rquickBody/ are mutually recursive, \LH/ requires that the user defines a termination metric for both functions.
We can use the tuple \C!(len l, 1)! for \rquick/ and \C!(len l, 0)! for \rquickBody/.
In \C!rquick!, the entry \C!len l! stays the same while \C!1! decreases to \C!0!.
Furthermore \C!len l! decreases in every call made from \rquickBody/, since the lengths of \C!sm! and \C!bg! sum up to \C!len (remIdx l i)!, which is equal to \C!len l - 1!.

The function \C!idx l i! returns the element at zero-based index \C!i! in list \C!l!, where the refinement type of \C!i! must satisfy \C!{i:#N | i < len l}!.
Recalling the refinement type annotation of \uniformBind/, it holds that \C!0 <= i <= len l - 1!.
Similar to \C!idx!, \C!remIdx l i! removes the \C!i!-th index of the list and we have \C!len (remIdx l i) + 1 == len l!.
The function \partition/ \C!x l! splits the list into one list with elements that are smaller than \C!x! and one list with elements that are bigger than \C!x!.
Moreover, the sum of their lengths equals the length of l and the union of their elements equals the element set of l.
In the listing, $\cup$ is an abbreviation for the \C!Set_cup! operation provided by \LH/ that is used to express the set union operation.
The quicksort procedure is then called once on the list of smaller values \C!sm! and once on the list of bigger values \C!bg!.
To count all the comparisons done by \partition/, the \tick/ function is used to incur costs of \C!len l - 1!, as the pivot is compared with all other list elements.

From the refinement type annotations we can deduce the following types for \C!rquick sm! and \C!rquick bg!:
\begin{lstlisting}[style=haskell]
(~\rquick/~ sm) :: ProperDist ({r:IncrList {v:#a | v < idx l i} | set r = set sm})
(~\rquick/~ bg) :: ProperDist ({r:IncrList {v:#a | idx l i < v} | set r = set bg})
\end{lstlisting}

To complete the implementation of randomised quicksort, the \combine/ function of the probability monad can be employed together with the \merge/ helper function.
For \merge/, we again exploit abstract refinement types to obtain that the resulting list is an increasing list.
Since both distributions of the input of \combine/ are proper, we automatically have that the resulting distribution is proper as well.
Moreover, each list occurring in \C!combine (rquick sm) (rquick bg) (merge (idx l i))! is of type \C!IncrList #a!.
It only remains to show that the elements of each of these lists is equal to the elements of \C!l!.
For each list \C!r! in the result distribution, it is inferred that \C!set r! is equal to \C[breaklines]!(Set_sng (idx l i)) $\cup$ set sm $\cup$ set bg!.
Since \C!sm! and \C!bg! result from a call to \partition/, it can be obtained that \C!set (remIdx l i)! equals \C!set sm $\cup$ set bg!.
Additionally, because \C!set l! is equal to \C!set (remIdx l i) $\cup$ (Set_sng (idx l i))! for lists without duplicate elements, the functional correctness for \rquick/ is established.

\paragraph{Expected Cost}
We give a proof in \LH/ that the expected amount of comparisons done by randomised quicksort is equal to $2(n+1)H_n - 4n$ for lists of length $n$ that contain unique elements.
This choice was also made by Cormen \emph{et al.} \cite{Cormen:2009}, Knuth \cite{knuth1998art} and Eberl \emph{et al.} \cite{EberlHN20}.
In our formal proof development we broadly follow the steps of Cicho\'{n} \cite{cichon-quick}.

The proof verifying the exact number of expected steps of randomised quicksort is separated into two parts.
First, we show via an extrinsic theorem that \C!$\HEcostOpen$rquick l$]$ = q_rec (len l)!, where the quicksort recurrence \C!q_rec! is defined as follows:
\begin{lstlisting}[style=haskell]
{-@ q_rec :: #N -> #R @-}
q_rec 0 = 0
q_rec n = n - 1 + (1 / n) * $\HfinSum{0}{n-1}$ (\i -> q_rec i + q_rec (n-1-i))
\end{lstlisting}
Afterwards, derive the closed form solution of \C!q_rec n!.

In the following, describe the first theorem that establishes \C!$\HEcostOpen$rquick l$]$ = q_rec (len l)!.
Since the proof goes by induction on the length of the list \C!l!, we can make use of the induction hypothesis for calls of \rquick/ that use lists of smaller length than \C!l!.
The base case \C!l = []! is trivial, since \dunit/ incurs no costs.
Thus, assume that \C!l! is non-empty.
Using the refinement annotations of \uniformBind/, we can verify via an extrinsic theorem that \C!$\HEcostOpen$rquick l$]$!, is given by
\footnotesize
\begin{align*}
(1 / \mathtt{len\ l}) \cdot {\textstyle \HfinSum{0}{\mathtt{len}\ \mathtt{l} - 1}}\ (\lambda \mathtt{i} \to \mathtt{len}\ \mathtt{l} - 1 &+ \HEcostOnly[ \rquick/\ (\mathtt{fst}\ (\mathtt{partition}\ (\mathtt{idx}\ \mathtt{l}\ \mathtt{i})\ (\mathtt{remIdx}\ \mathtt{l}\ \mathtt{i})))] \\
                                                                                                                                         &+ \HEcostOnly[ \rquick/\ (\mathtt{snd}\ (\mathtt{partition}\ (\mathtt{idx}\ \mathtt{l}\ \mathtt{i})\ (\mathtt{remIdx}\ \mathtt{l}\ \mathtt{i})))]).
\end{align*}\normalsize

We now use several of the theorems provided by our finite summation module (see \Cref{fig:finSum-theorems}).
By applying theorem \thmSumLinear/, we can extract the expression \C!len l - 1! into its own summation \C!$\HfinSum{0}{\mathtt{len}\ \mathtt{l} - 1}$ (\i -> len l - 1)!, which can be rewritten into \C!(len l) * (len l - 1)! with the help of \thmSumConstant/.
Additionally, we can appeal to the induction hypothesis as the length of both lists returned by \partition/ is smaller than \C!len l!.
Thus, the simplified expression for \C!$\HEcostOpen$rquick l$]$! is
\footnotesize
\begin{align*}
\mathtt{len\ l} - 1 + (1 / \mathtt{len\ l}) \cdot {\textstyle \HfinSum{0}{\mathtt{len}\ \mathtt{l} - 1}}\ (\lambda \mathtt{i} \to &\ \mathtt{q\_rec}\ (\mathtt{len}\ (\mathtt{fst}\ (\mathtt{partition}\ (\mathtt{idx}\ \mathtt{l}\ \mathtt{i})\ (\mathtt{remIdx}\ \mathtt{l}\ \mathtt{i})))) \\
                                                                                                                                + &\ \mathtt{q\_rec}\ (\mathtt{len}\ (\mathtt{snd}\ (\mathtt{partition}\ (\mathtt{idx}\ \mathtt{l}\ \mathtt{i})\ (\mathtt{remIdx}\ \mathtt{l}\ \mathtt{i})))))
\end{align*}\normalsize

In order to further simplify the above formula, we want to get rid of the calls to \partition/.
This is achieved through the notion of \emph{rank} as follows.
Given a list $\ell$ and a value $x$ occurring therein, the rank of $x$ is given by the number of elements of $\ell$ that are smaller than $x$.
For lists $\ell$ with distinct elements, the rank of each element is also unique and contained in $\{0,\ldots, |\ell|-1\}$.
Hence, a list element can be identified either by its rank or its index.
So instead of choosing an index it is equivalent to choose a rank uniformly at random when selecting the pivot element.
The \C!rank! function defined in \Cref{fig:rank} allows us to formalise this fact.

\begin{figure}[t]
\begin{lstlisting}[style=haskell,belowskip=0pt,numbers=left]
{-@ rank :: Ord #a => l:[#a] -> x:#a ->
      {r:#N | (r <= len l) && (x $\listIn/$ l => (r < len l))
          && (r = len (fst (partition x l))) ~\label{lst:rank:ref1}~
          && ((not (x $\listIn/$ l)) => ((len l) - r = len (snd (partition x l))))} @-} ~\label{lst:rank:ref2}~
rank [] _ = 0
rank (x:xs) y = if y > x then 1 + rank xs y else if y == x then rank xs y else rank xs y
\end{lstlisting}
\caption{Rank}
\label{fig:rank}
\end{figure}

By exploiting the refinements expressing the relationship between \C!rank! and \partition/ (see lines~\ref{lst:rank:ref1} and \ref{lst:rank:ref2} in \Cref{fig:rank}) together with a theorem establishing that \C!rank (remIdx l i) (idx l i)! is equal to \C!rank l (idx l i)!, we can replace the calls to \partition/ as given below:
\footnotesize
\[
\mathtt{len\ l} - 1 + (1 / \mathtt{len\ l}) \cdot {\textstyle \HfinSum{0}{\mathtt{len}\ \mathtt{l} - 1}}\ (\lambda \mathtt{i} \to \mathtt{q\_rec}\ (\mathtt{rank}\ \mathtt{l}\ (\mathtt{idx}\ \mathtt{l}\ \mathtt{i})) + \mathtt{q\_rec}\ (\mathtt{len}\ \mathtt{l} - 1 - \mathtt{rank}\ \mathtt{l}\ (\mathtt{idx}\ \mathtt{l}\ \mathtt{i})))
\]\normalsize

We also get rid of the calls to \C!rank! by applying the theorem \thmSumPermute/, which allows to permute the sum indices.
The permutation is defined using the two functions given below.
\begin{lstlisting}[style=haskell]
{-@ qsF :: Ord #a => l:NoDupList #a -> {i:#N | i < len l} -> {rk:#N | rk < len l} @-}
qsF l i = rank l (idx l i)

{-@ qsG :: Ord #a => l:NoDupList #a -> {rk:#N | rk < len l} -> {idx:#N | idx < len l} @-}
qsG l rk = idxOf l (idx (sort l) rk)
\end{lstlisting}

The function \C!qsF l i! computes the rank of the list element at index \C!i!, whereas \C!qsG l rk! evalutes to the index of the element in \C!l! whose rank is equal to \C!rk!.
With the help of a theorem certifying the properties that \C!qsF l (qsG l x) == x! and \C!qsG l (qsF l x) == x!, we can apply \thmSumPermute/ in order to permute the list indices using the function \C!qsG!.
Because of the property validating that these functions are inverses of each other, we finally have:
\footnotesize
\begin{align*}
\HEcostOnly[\rquick/\ \mathtt{l}] &= \mathtt{len\ l} - 1 + (1 / \mathtt{len\ l}) \cdot {\textstyle \HfinSum{0}{\mathtt{len}\ \mathtt{l} - 1}}\ (\lambda \mathtt{i} \to \mathtt{q\_rec}\ \mathtt{i} + \mathtt{q\_rec}\ (\mathtt{len}\ \mathtt{l} - 1 - \mathtt{i})) \\
                                  &= \mathtt{q\_rec}\ (\mathtt{len\ l}).
\end{align*}\normalsize

The second step for deriving a closed form for \C!$\HEcostOpen$rquick l$]$! is to solve the recurrence encoded by \C!q_rec!.
First, we show by an extrinsic theorem that \C!q_rec n == q_rec' n! for all natural numbers \C!n!, where
\begin{lstlisting}[style=haskell]
{-@ q_rec' :: #N -> #R @-}
q_rec' 0 = 0
q_rec' n = n - 1 + (2 / n) * $\HfinSum{0}{n-1}$ (\i -> q_rec' i)
\end{lstlisting}

The proof is simple by applying the summation theorems \thmSumLinear/ and \thmSumReverse/.
Then, the recurrence \C!q_rec'! can be rewritten into a single summation as given below.

\begin{lstlisting}[style=haskell]
{-@ reflect sum_term @-}
{-@ sum_term :: {k:#Z | k >= 1} -> #Rnneg @-}
sum_term k = (k-1) / (k * (k + 1))

{-@ thm_q_rec_sum :: n:#N -> {q_rec' n = (n+1) * 2 * $\HfinSum{1}{n}$ sum_term} @-}
thm_q_rec_sum 0 = ()
thm_q_rec_sum 1 = ()
thm_q_rec_sum n = $\HfinSum{1}{n-1}$ q_rec' |> $\HfinSum{1}{n}$ sum_term |> thm_q_rec_sum (n-1)
\end{lstlisting}

Note that in the inductive proof of \C!thm_q_rec_sum! the cases for \C!n = 0! and \C!n = 1! are discharged automatically.
For \C!n >= 2!, the expressions involving the finite sums are added such that the refinement type properties allowing their unfoldings can be exploited by the SMT solver together with an application of the induction hypothesis.
Hence, \C!q_rec n! is equal to the sum $\sum_{k=1}^{n} \frac{k-1}{k(k+1)}$.
Since $\frac{k-1}{k(k+1)}$ can be written as $\frac{2}{k+1}-\frac{1}{k}$, we can employ the summation theorem \thmSumLinear/ and an encoding of harmonic numbers in \LH/
defined by the equations \C!harmonic 0 = 0! and \C!harmonic n = (1/n) + harmonic (n-1)! to show that \C!$\HfinSum{1}{n}$ sum_term! is equal to \C!(2/(n+1) + harmonic n - 2)!.
Moreover, because \C!q_rec' n! equals \C!(n+1) * 2 * $\HfinSum{1}{n}$ sum_term!, we can deduce the following equality:

\footnotesize
\begin{center}
\ttfamily  q\_rec' n = 2*(n+1) * (harmonic n) - 4*n\normalfont.
\end{center}\normalsize

Combining this result with the theorems establishing that
\C!$\HEcostOpen$rquick l$]$! equals \C!q_rec (len l)! and \C!q_rec n! is equal to \C!q_rec' n!, we have derived the final result
that the expected number of comparisons performed by randomised quicksort is given by \C!2*(n+1) * (harmonic n) - 4*n!, representing ${2(n+1)H_n - 4n}$.

\end{document}